\documentclass{article}
\usepackage{authblk}
\usepackage[utf8]{inputenc}
\usepackage{url}
\usepackage[version=4]{mhchem}
\usepackage{multirow}
\usepackage{lipsum}
\usepackage{stackrel}
\usepackage{listings}
\usepackage{amssymb}
\usepackage{nccmath}
\usepackage{a4wide}
\usepackage{mathtools}
\usepackage{amsmath,amsfonts,amsthm,amssymb}
\usepackage{bbm, dsfont}
\usepackage[justification=justified, singlelinecheck=false]{caption}
\usepackage{graphicx}
\usepackage{enumitem}
\usepackage{empheq,etoolbox}
\patchcmd{\subequations}
  {\theparentequation\alph{equation}}
  {\theparentequation.\alph{equation}}
  {}{}
\usepackage[justification=centering]{caption}
\usepackage{algorithmicx}
\usepackage{algorithm}
\usepackage{algpseudocode}
\usepackage{stackengine}
\newcommand\dhookrightarrow{\mathrel{%
  \ensurestackMath{\stackanchor[.1ex]{\hookrightarrow}{\hookrightarrow}}
}}

\usepackage[title]{appendix}
\usepackage[scaled=.90]{helvet}
\usepackage{courier}
\usepackage{ae}
\usepackage[T1]{fontenc}
\usepackage{subfigure}
\usepackage{graphicx,color}
\usepackage{xcolor}
\usepackage{here} 
\usepackage[colorlinks=false, pdfborder={0 0 0}]{hyperref}
\usepackage[symbol]{footmisc}

\DeclareMathOperator{\Tr}{Tr}
\usepackage{mdwlist}
\usepackage{tcolorbox}
\usepackage{fancybox}
\usepackage{framed}
\usepackage{verbatim}

\definecolor{darkblue}{rgb}{0,0,0.8}
\definecolor{darkgreen}{rgb}{0,0.8,0}

\definecolor{magenta}{rgb}{0.5,0,0.5}

\newcommand{\mathleft}{\@fleqntrue\@mathmargin0pt}

\newtheorem{theorem}{Theorem}[section]
\newtheorem{definition}{Definition}[section]

\newtheorem{assumption}{Assumption}[section]

\newtheorem{proposition}{Proposition}[section]
\newtheorem{remark}{Remark}[section]

\providecommand{\keywords}[1]
{
  \small	
  \textbf{\textit{Keywords---}} #1
}

\usepackage[maxbibnames=99]{biblatex}

\begin{document}
\title{A stochastic hierarchical model for low grade glioma evolution} 
\author[1,2]{Evelyn Buckwar}
\author[3]{Martina Conte}
\author[1]{Amira Meddah}
\affil[1]{\centerline{\small Institute of Stochastics, Johannes Kepler University Linz,} \newline \centerline{\small Altenberger Straße 69, 4040 Linz, Austria}}
\affil[2]{\centerline{\small Centre for Mathematical Sciences, Lund University, 221 00 Lund, Sweden}}
\affil[3]{\centerline{\small Department of Mathematical Sciences "G. L. Lagrange", Politecnico di Torino} \newline \centerline{\small Corso Duca degli Abruzzi 24 - 10129 Torino, Italy}}
\date{\today}                     
\setcounter{Maxaffil}{0}
\renewcommand\Affilfont{\itshape\small}
\maketitle
\abstract{A stochastic hierarchical model for the evolution of low grade gliomas is proposed. Starting with the description of cell motion using piecewise diffusion Markov processes (PDifMPs) at the cellular level, we derive an equation for the density of the transition probability of this Markov process using the generalised Fokker-Planck equation. Then a macroscopic model is derived via parabolic limit and Hilbert expansions in the moment equations. After setting up the model, we perform several numerical tests to study the role of the local characteristics and the extended generator of the PDifMP in the process of tumour progression. The main aim focuses on understanding how the variations of the jump rate function of this process at the microscopic scale and the diffusion coefficient at the macroscopic scale are related to the diffusive behaviour of the glioma cells and to the onset of malignancy, i.e., the transition from low-grade to high-grade gliomas.}\\\vspace{0.05cm}

 
 \section{Introduction} 
 
 Gliomas are the most common type of primary brain tumours, accounting for $78\%$ of all malignant brain neoplasia \cite{Neur_academy}. They originate from mutations of the glial cells in the central nervous system and are classified by the World Health Organisation (WHO) into four grades according to the degree of malignancy (see \cite{wesseling20182016} for a more detailed description). In this work, we mainly focus on the low grade gliomas (LGGs), which are a class of rarely curable diseases, often resulting in the premature death of the patient. Since in the last years some medical interventions have shown to improve the median survival time of the patients, the study of this class of tumour has become of great importance for the clinicians.\\
\indent The development, growth, and invasion of gliomas in the brain is a very complex phenomenon, involving many interrelated processes over a wide range of spatial and temporal scales. As such, often the individual cell behaviours and the intracellular dynamics described at a microscopic scale are manifested by functional changes in the cellular and tissue level phenomena. Therefore, this multiscale nature of glioma evolution requires modelling techniques that are able to deal with different levels of description.\\
\indent The first mathematical models for the study of brain tumours started to emerge in the early 1980s (see \cite{duchting1980spread, duchting1981three, duchting1985recent,tracqui1995mathematical, tracqui1995passive} for further details). Since then, the mathematical modelling of glioma evolution has evolved considerably and several different approaches have been proposed, going from discrete or hybrid microscopic models to macroscopic and multiscale frameworks. Discrete models at the microscopic scale, also called agent-based models, have been used to describe the dynamics of individual cells moving on a lattice (for some examples we refer the reader to \cite{wang2015simulating, metzcar2019review, hatzikirou2012go}, or, specifically, to  \cite{aubert2006cellular} for cellular automata models and \cite{gao2013acute} for cellular Potts models). Further, stochastic discrete models for cell motion have also been proposed, e.g. describing $2$D persistent random walk or $3$D anomalous diffusion \cite{dunn1987unified,luzhansky2018anomalously,audoin2022tumor,scott2021mathematical}. In particular, recently in \cite{scott2021mathematical}, the authors have presented the analysis of $3$D cell tracking data, based on a persistent random walk model adapted into the context of glioma cell migration. At the macroscopic scale, several phenomenological models for glioma evolution stated in the form of reaction-diffusion-advection equations have been proposed and studied \cite{swanson2003virtual, harpold2007evolution, tracqui1995mathematical, swanson2000quantitative}, also including patient-specific data (e.g. in the form of diffusion tensor imaging (DTI) information). This has allowed for a comparison between the real and the virtual tumour evolution \cite{jbabdi2005simulation, konukoglu2010extrapolating, clatz2005realistic, mosayebi2012tumor}. Concerning multiscale models, a broad and rich literature has been developed for the integration of microscopic and macroscopic dynamics (for some examples see  \cite{hillen2006m5, hillen2013transport, bellomo2012asymptotic, painter2013mathematical,engwer2015glioma, engwer2016effective, lorenz2014class, kelkel2012multiscale,engwer2016multiscale}). In particular, in \cite{engwer2015glioma}, a more detailed description of the migration process of individual cells, involving the dynamics of cell receptors and the interaction with the tumour microenvironment, is discussed. \\
\indent A key aspect of modelling tumour evolution concerns cell movement, which is based on a combination of complex processes involving motility and migration: motility refers to the random movement from one location to another, while migration involves also the interactions between cells and the microenvironment \cite{motility_migration}.\\
\indent The first description of particle movement, which uses a stochastic Markov process combining deterministic ordinary differential equations (ODEs) for the continuous movement with Poisson-like jumps for the random change of direction, was introduced in 1974 by Stroock \cite{stroock1974some} on the basis of the biological observations illustrated in \cite{adler1966chemotaxis}. The concept of piecewise deterministic Markov processes (PDMPs) was introduced in 1984 in \cite{Davis}. An extension of \cite{Davis} was then provided in \cite{bujorianu2004general, bujorianu2006toward}, where the authors developed the extended generator and the differential formula for piecewise diffusion Markov processes (PDifMPs), showing that all the classes of proposed stochastic hybrid processes can be seen as a special case of their  concept of a general stochastic hybrid system (GSHS).  Further, in \cite{blom1988piecewise} a general class of continuous-time stochastic hybrid systems in which the continuous flow is the solution flow of a stochastic differential equation (SDE) was presented. These processes have been widely applied in different contexts, e.g. for  interacting particle systems \cite{blom2018interacting}, air traffic management \cite{bujorianu2005toward}, or gene network \cite{nankep2018modelisation}) and especially in biological modelling (for some examples, see \cite{uatay2019multiscale, fontbona2010quantitative,cloez2017probabilistic, pakdaman2010fluid,genadot2014multiscale,riedler2011spatio,Debussche2012,TyranKaminska2017}). However, it seems that the use of PDifMPs in the context of tumour growth, motility, and migration has not yet been investigated. In this article we extend the description of cell movement based on velocity jump processes with the use of PDifMPs in the context of glioma progression. In particular, we build a multiscale model, starting with a contact-mediated description of cell motion on the microscopic scale using PDifMPs. We use the extended generator for such processes to derive a generalised Fokker-Planck equation, including the description of the tumour-microenvironment interactions. The solution of this equation provides the joint density of the transition probabilities of this Markov process for all the involved variables. As the variables involved in these interactions are fast-acting compared to the macroscopic scale, we make use of a scale separation variable and the Hilbert expansion method to derive the corresponding macroscopic scale equation for the time and space variables (for a more general discussion of multiscale modelling and moment closure techniques, we refer the reader to \cite{Vicecontieetal2021multiscale,KuehnMomentClos2016,hunt2018dti}).\\
\indent The paper is organised as follows. Section \ref{section2} contains a brief introduction to PDifMPs. In Section \ref{section3}, we derive a stochastic multiscale model for glioma progression. Numerical simulations in a $2$D scenario for the resulting macroscopic equation for the tumour cell density are presented in Section \ref{section4}, including several studies on the effect of parameter variations. Finally, in Section \ref{section5}, we review our results and discuss further directions of research.
 
\section{Preliminaries on PDifMPs}
\label{section2}
\subsection{Definition and notation}
In this section, we provide a brief introduction to PDifMPs and the construction of their paths. We refer the reader to \cite{bujorianu2006toward} and \cite{ nankep2018modelisation} for a general description of stochastic hybrid systems.\\
\indent Let $(\Omega, \mathcal{F}, (\mathcal{F})_{t \geq 0}, \mathbb{P})$ be a filtered probability space and $(W_t)_{t\in [0,T]}$ an {$m$-dimensional} standard Wiener process, with $m \in \mathbb{N}$ and $T>0$. We consider the PDifMP defined by ${(U_t)_{t\in [0,T]}:=\{ U(t,\omega), t\in [0,T], \omega \in \Omega \}}$. It consists of two different components, i.e., ${U_t=(S_t,V_t)}$ with values in $E=E_1\times \mathbf{V}$. In particular, $E_1 \subset \mathbb{R}^{d_1}$ and $\mathbf{V} \subset \mathbb{R}^{d_2}$, with $d_1,d_2 \in \mathbb{N}$ and $E$ endowed with the Borel algebra $\mathcal{B}(E)$. The closure of the set $E$ is denoted by $\Bar{E}$, while $\partial E$ stands for its boundary.\\
\indent For the couple of non-exploding processes $(S_t,V_t)$, we assume that the first stochastic component $(S_t)_{t\in [0,T]}$ possesses continuous paths in $E_1$ and the second component $(V_t)_{t\in [0,T]}$ is a jump process with right continuous paths and piecewise constant values in $\mathbf{V}$. The times $(T_i)_{i\in \mathbb{N}}$ at which the second component jumps form a sequence of randomly distributed grid points in $[0,T]$.\\
The motion of the PDifMP $(U_t)_{t\in [0,T]}$ on $(E,\mathcal{B}(E))$ is defined by its characteristic triple $(\phi, \lambda, \mathcal{Q})$ as follows:
\begin{itemize}
	\item $\phi: [0,T] \times E \rightarrow E_1$, $(t,u)\mapsto \phi (t,u)$, is the stochastic flow of the continuous first component of $(U_t)_{t\in [0,T]}$. Starting at $T_0=0$ with initial value ${u_0=(s_0,v_0)\in E}$, the process $\phi(t,u)$  represents the solution of a sequence of SDEs over the consecutive intervals $[T_i, T_{i+1})$ of random length. At each random point $T_i\in [0,T]$, $i\geq 1$, there are newly updated initial values $u_i=(s_i,v_i)\in E$, where $s_i$ serves as the initial value and $v_i$ as a parameter in the following SDE defined on the interval $[T_i, T_{i+1})$:
	
	\begin{equation}
		\label{sys_1}
		\left\{
		\begin{array}{ll}
			d\phi(t,u_i)=b(\phi(t,u_i),v_i)dt+\sigma(\phi(t,u_i),v_i)dW_t, \qquad t \in [T_i, T_{i+1}),\\[0.2cm]
			\phi(T_i,u_i)=s_i.
		\end{array}
		\right. 
	\end{equation}
	At the end point $T_{i+1}$ of each interval, $s_{i+1}$ is set to the current value of $\phi(\,\cdot\,, u_i)$ to ensure the continuity of the path. Further, a new value $v_{i+1}$ is chosen as fixed parameter for the next interval according to the jump mechanism described below. We define also the function $b$ with values in $\mathbb{R}^{d_1}$, which represents a family of drift coefficients, and the ${d_1}\times m$ matrix $\sigma$ with real coefficients.
	\begin{assumption}
		\label{existence}
		We assume that $b:E\rightarrow \mathbb{R}^{d_1}$ and $\sigma:E\rightarrow \mathbb{R}^{{d_1}\times m}$ are linearly bounded and globally Lipschitz continuous for all $s\in E_1$. 
	\end{assumption}
	For any $v_i\in \mathbf{V}$, this assumption ensures the existence and uniqueness of the solution to (\ref{sys_1}) (see Theorem 5.2.1 in \cite{oksendal2013stochastic}). Moreover, the stochastic flow satisfies the semi-group property, i.e.,
	
	\begin{equation*}
		\phi_{v_i}(t+\delta,\,\cdot\,)=\phi_{v_i}(t,\phi_{v_i}(\delta,\,\cdot\,))\,,\qquad \forall t,\delta \in [0,T]\,.
	\end{equation*}
	
	\item $\lambda: E \rightarrow \mathbb{R}_{+}$ is the jump rate, i.e, it determines the frequency at which the second component of $(U_t)_{t\in [0,T]}$ jumps.
	\item $\mathcal{Q}:(E,\mathcal{B}(E))\rightarrow [0,1]$ is the transition kernel that determines the new values of the second component after a jump occurs. For all $u\in E$, it satisfies $\mathcal{Q}(u,\{u\})=0$, meaning that the process cannot have a no-move jump.  
\end{itemize}
Moreover, for all $t\in [T_i, T]$, $i\geq 0$, we define the survival function of the inter-jump times as

\begin{equation}
	\label{gen_surv}
	\mathcal{S}(t,u_i):= \exp \left(-\int_{T_i}^t \lambda(\phi(\delta,u_i),v_i)d\delta \right), \qquad u_i\in E.
\end{equation}
This function states that there is no jump in the time interval $[T_i,t)$ conditional on the process being in the initial state $u_i$. Let $\mathcal{U}$ be a uniformly distributed random variable on $[0,1]$, thus $\zeta:[0,1]\times E \rightarrow \mathbb{R}_{+}$ is the generalised inverse of $\mathcal{S}(t,u_i)$ defined by

\begin{equation*}
	\zeta(\mathcal{U},u_i)=\inf\{t\geq 0 \, ;\, \mathcal{S}(t,u_i)\leq \mathcal{U}\}.
\end{equation*}
\begin{assumption}
	\label{assump_lambda}
Let	$\lambda:E\rightarrow \mathbb{R}_{+}$ be a measurable function such that $\forall u_i\in E$ and $T>0$
	
	\begin{equation}
		\int_0^T \lambda(\phi(t,u_i),v_i) < \infty \hspace{1cm} \text{and} \hspace{1cm} \int_0^{\infty} \lambda(\phi(t,u_i),v_i) = \infty. 
	\end{equation}
\end{assumption}
\noindent Moreover, there exists a measurable function $\psi:[0,1]\times E \rightarrow E$ such that for $u_i\in E$ and $A\in \mathcal{B}(E)$

\begin{equation*}
	\mathbb{P}(\psi(\mathcal{U},u_i)\in A)=\mathcal{Q}(u_i,A).
\end{equation*}
$\psi$ represents the generalised inverse function of $\mathcal{Q}$. For a fixed $t$, $\psi(\mathcal{U}(\omega),U(\omega))$ is a random variable describing the post-jump locations of the second component of $U$.
\begin{assumption}
	For all $A\in \mathcal{B}(E)$, $\mathcal{Q}(\,\cdot\,,A)$ is measurable, while for all $u\in \Bar{E}$ the function $\mathcal{Q}(u,\,\cdot\,)$ is a probability measure.
\end{assumption}
\noindent Summarising, the first component of the triple $(\phi,\lambda, \mathcal{Q})$ describes the continuous evolution of the trajectories of the process $(U_t)_{t\in [0,T]}$ between jumps in time intervals defined by the survival function $\mathcal{S}$, while the couple $(\lambda, \mathcal{Q})$ yield the jump mechanism. All three components of $(\phi,\lambda, \mathcal{Q})$ are coupled. 

\subsection{Construction}
\label{const}
From the local characteristics $(\phi,\lambda, \mathcal{Q})$, it is possible to iteratively construct the sample path $U_t$ as follows.
Let $(\mathcal{U}_n)_{n\geq 1}$ be a sequence of iid random variables with uniform distribution on $[0,1]$ and  $u_0=(s_0,v_0)\in E$ the initial value of (\ref{sys_1}) at $T_0=0$, such that $u_0$ can be either an $\mathcal{F}_0$-measurable random variable (independent from the Wiener process) or a deterministic constant, for some $\omega \in \Omega$. We apply the survival function $\mathcal{S}(t,u_0)$ defined in (\ref{gen_surv}) and use its generalised inverse $\zeta$ with the first element $\mathcal{U}_1$ to determine ${T_1=\zeta(\mathcal{U}_1,u_0)}$, i.e., the first jump time of the second component of $U_t$. We then define the sample path $U_t$ up to the first jump time as

\begin{equation*}
	\left\{
	\begin{array}{ll}
		U_t=\phi(t, u_0) \qquad\qquad\qquad\qquad \text{for}~ 0\leq t<T_1, \\[0.2cm]
		U_{T_1}=\psi\left(\mathcal{U}_2,\big(\phi(T_1,u_0),v_0\big)\right).
	\end{array}
	\right. 
\end{equation*}
The trajectory of $U_t$ follows the stochastic flow $\phi$ given in (\ref{sys_1}) starting from $U_0=u_0$ until a first jump occurs at the random time $t=T_1$. The post-jump state $U_{T_1}$ is determined through the measurable function $\psi$. For all $A\in \mathcal{B}(E)$, the distribution of $\psi(\mathcal{U}_2,u_0)$ is given by

\begin{equation}
 	\mathbb{P}(V_{T_1}\in A\vert t=T_1, S_0=s_0)=\mathcal{Q}\left((\phi(\tau_1,u_0),v_0),A\right),
\end{equation}
where $\tau_1$ is the waiting time until the first jump occurs, i.e. $\tau_1=T_1$.\\
Restarting the process from the post-jump location $U_{T_1}$, we define

\begin{equation*}
	\tau_2=\zeta(\mathcal{U}_3,U_{T_1})
\end{equation*}
the next waiting time before a jump occurs from the survival function (\ref{gen_surv}). In this way, we find the next jump time $T_2=T_1+\tau_2$.\\
Consequently, the state of the process in the interval $[T_1,T_2)$ is given by

\begin{equation*}
	\left\{
	\begin{array}{ll}
		U_t=\phi(t-T_1, U_{T_1}) \qquad\qquad\qquad\qquad\qquad \text{for}\quad T_1\leq t<T_2, \\[0.2cm]
		U_{T_2}=\psi\left(\mathcal{U}_3,\big(\phi(\tau_2,(U_{T_1},v_0)),v_0\big)\right).
	\end{array}
	\right. 
\end{equation*}
We proceed recursively to obtain a sequence of jump times $(T_i)_{i\geq 1}$, 

\begin{equation*}
	{T_i=T_{i-1}+\zeta(\mathcal{U}_{2i-1},U_{T_{i-1}})} \qquad \forall i\geq 1,
\end{equation*}
such that the generic sample path of $U_t$, for $t\in [T_i, T_{i+1})$, is defined accordingly by

\begin{equation*}
	\left\{
	\begin{array}{ll}
		U_t=\phi(t-T_i, U_{T_i}) \qquad\qquad\qquad\qquad\qquad\qquad \text{for}\quad T_i\leq t<T_{i+1}, \\[0.2cm]
		U_{T_{i+1}}=\psi\left(\mathcal{U}_{2i+2},\big(\phi(\tau_{i+1},(U_{T_i},v_{i+1})),v_{i+1}\big)\right).
	\end{array}
	\right. 
\end{equation*}
The number of jump times that occur between $0$ and $t$ is denoted by 

\begin{equation*}
	N_t=\sum_{i\geq 1} \mathbbm{1}_{(T_i\leq t)}.
\end{equation*}
\begin{assumption}
	\label{nnexp}
	For all $t>0$ and for every starting point $u_i\in E$, $\mathbb{E}[N_t\vert u=u_i]<\infty$.
\end{assumption}
\noindent This assumption ensures the non-explosion of the process $U_t$. Under the Assumptions \ref{existence}-\ref{nnexp} the piecewise diffusion process can be constructed as a strong c\`adl\`ag Markov process (see \cite{bujorianu2006toward} for further details), called then a Piecewise Diffusion Markov Process (PDifMP).
\subsection{Extended generator of the PDifMP}
\label{2E}
The notion of infinitesimal generator is an extremely important tool for the study of Markov processes \cite{bielecki2006extended, Davis}. In the following, we adopt the definition in \cite{nankep2018modelisation, bujorianu2006toward}, and, for the reader's convenience, we recall the theorem that fully characterised the extended generator (see \cite{bielecki2006extended} and references therein for further details about the difference between extended and classic generators).
\begin{theorem}
	Let $U_t$ be a PDifMP with characteristics $(\phi,\lambda,\mathcal{Q})$. The domain $\mathcal{D}(\mathcal{A})$ of the extended generator $\mathcal{A}$ consists of all bounded, measurable functions $f$ on $E \cup \partial E$ satisfying:
	\begin{enumerate}
		\item $f: \bar{E} \rightarrow \mathbb{R}$  $\mathcal{B}$-measurable such that $s\mapsto f(s,v)$ is a.e $C^2(\bar{E})$,
		\item
		
		$$
		f(u)=\int_{E} f(y) \mathcal{Q}(u, dy), \:  u \in \partial E,
		$$
		\item  $B f \in L_{1}^{\text{loc }}(p)$  where 
	
		$$
		B f(u, t, \omega):=f(u)-f\left(u_{t-}(\omega)\right) .
		$$
	\end{enumerate}
	Then, for $f\in \mathcal{D}(\mathcal{A})$, $u=(s,v)\in E$, the extended generator $\mathcal{A}f$ is given by 
	
	\begin{equation}
		\label{eqn:Extended_gen}
		\mathcal{A}f(s,v)=\mathcal{A}_{\text{dif}}f(s,v)+\lambda(s,v)\int_{E}(f(s,\xi)-f(s,v))\mathcal{Q}((s,v),d\xi), 
	\end{equation}
	where
	
	\begin{equation} \label{eq1}
		\begin{split}
			\mathcal{A}_{\text{dif}}f(s,v)&:=\sum_{i=1}^{d_1} b_i(s,v)\partial_i f(s,v)+ \frac{1}{2} \sum_{i,j=1}^{d_1}(\sigma \sigma^T)_{ij}(s,v)\partial_i \partial_j f(s,v),\\
			& = \nabla_sf(s,v)\cdot b(s,v)+\frac{1}{2}\Tr[(\sigma \sigma^T)(s,v)(\nabla_s\nabla_s^T)f(s,v)],
		\end{split}
	\end{equation} 
	for $s=(s_1,\ldots,s_{d_1})$. Here, $\nabla_sf(s,v)\cdot b(s,v)$ is the inner product in $\mathbb{R}^{d_1}$, $\sigma^T$ is the transpose matrix of $\sigma$, $\nabla_s^T$ is the transpose operator of $\nabla_s$.
\end{theorem}
\noindent We refer to \cite{bujorianu2006toward} for the definition of $L_{1}^{\text{loc }}(p)$ and the proof of this theorem.
\subsection{Generalised Fokker-Planck equation}
\label{GFPE}
The adjoint of the generator is used to derive the generalised Fokker-Planck equation, describing the time evolution of the probability distribution $g(t,s,v)$ of the process. The equation is given by

\begin{equation}
	\partial_t g(t,s,v)=\mathcal{A}_{\text{dif}}^{*}g(t,s,v)+\lambda(s,v)\int_{E}(g(s,\xi)-g(s,v))\mathcal{Q}((s,v),d\xi),
\end{equation}
where the adjoint operator of $\mathcal{A}_{\text{dif}}$ reads

\begin{equation}
	\mathcal{A}_{\text{dif}}^{*}g(t,s,v)=-\nabla_sg(t,s,v)\cdot b(s,v)+\frac{1}{2}\Tr\big[(\sigma \sigma^T)(s,v)(\nabla_s\nabla_s^T)g(t,s,v)\big].
\end{equation}
We refer to \cite{bect2007processus,Gardiner} for further details on the derivation of Fokker-Planck equations for general Markov processes.

\section{Application to tumour modelling}
\label{section3}
Gliomas can be considered as dynamical ecosystems where cells undergo constant changes due to many cellular processes, e.g. migration, proliferation, death, or creation of new blood vessels \cite{tamai2022tumour,ahir2020tumour}.
We focus on the process of cell movement, which is responsible for the global diffusive features that characterise glioma evolution. Cell movement can be divided into motility and migration. Motility refers to the random or spontaneous motion of cells from one location to another, while cell migration involves many interconnected biological aspects, such as environmental cues driving it. Thus, methods that take into consideration the stochastic nature of this phenomenon (i.e., motility) while accounting for environmental cues influencing it (i.e., migration) are important for providing a more complete understanding of the entire process.\\
\indent Following \cite{kelkel2012multiscale, hunt2018dti}, we model the process of cell movement under the influence of subcellular scale interactions, considering the effects of the amount of bound receptors located on the cell membrane. Specifically, we consider the role of integrins in this dynamics \cite{danen2013integrin, ellert2020integrin}. Referring to cell migration, we take into account the alignment of the tissue as a cue enhancing the efficiency of cell invasion \cite{vader2009strain,deutsch2007mathematical}, as cells tend to attach to the fiber and crawl along them, a phenomenon referred to as \textit{contact guidance}. However, since the direction that cells decide to follow remains random, there is a need to consider a stochastic description for the motility component.\\
\indent Inspired by particle movement models \cite{stroock1974some, othmer2000diffusion, othmer2002diffusion}, we propose piecewise diffusion Markov processes for the modelling of cell movement. In the context of persistent random motion, the continuous stochastic component of the PDifMP describes the contact guidance phenomenon, while its second component describes the random motility dependent on the velocity jump process.
This approach makes it possible to describe the cellular migratory response to environmental signals while keeping the random aspect of cell motility. Moreover, it also allows us to show how several well-established methods proposed in the literature (e.g. see \cite{othmer2000diffusion, othmer2002diffusion,hunt2018dti}) can be cast into a rigorous PDMP framework.
\subsection{Microscopic scale} 
\subsubsection{Interactions between cells and microenvironment}
\label{Sub3.1}
In order to migrate through the complex brain structure, glioma cells must adapt quickly to the physical characteristics of the environment. Their interactions with the extracellular matrix (ECM) \cite{frantz2010extracellular} are mediated by the binding between the integrins and the ECM fibrillar proteins. These bindings allow them to exert the forces necessary for them to migrate \cite{roth2013integrin, ellert2020integrin}. As these processes happen at a sub-cellular level, we describe the mechanism behind cell motion modelling the dynamics of the receptors on the tumour cell membrane.\\
\indent Let ${y(t) \in (0,1)}$ be the concentration of bound integrins and let us assume that the binding between integrins and tissue occurs in areas of highly aligned fibers \cite{engwer2015glioma}. The binding process can be described with the following general reaction

\begin{equation}
	\label{Reaction}
	\ce{Q +(R_0-y)
		<=>[\ce{k^{+}}][\ce{k^{-}}]y}  
\end{equation}
where $R_0$ defines the total number of cell surface receptors, $Q(x)$ the macroscopic volume fraction of tissue (including ECM and brain fibers), depending on the position $x\in \mathbf{X}\subset \mathbb{R}^3$, and $k^{+}$ and $k^{-}$ the rates of attachment and detachment between cell and tissue \cite{engwer2016multiscale, engwer2015glioma}. Within this framework, denoting by $x=x_0+vt$, we look at the path of a single cell moving from an initial position $x_0$ with velocity $v\in \mathbf{V} \subset \mathbb{R}^{3}$. $\mathbf{V}=\alpha\mathbb{S}^{2}$ is the closed set for cell velocities, where $\mathbb{S}^{2}$ denotes the unit sphere on $\mathbb{R}^{3}$ and $\alpha$ the mean speed of a tumour cell, which is assumed to be constant. Since we are interested in the interactions between cell surface receptors and the ECM, and this binding process takes place for fixed position $x$, we ignore any type of randomness resulting from the velocity change. The mass action kinetics for the concentration $y(t)$ is governed by the following ODE:

\begin{equation}
	\label{reac_eq}
	\frac{dy}{dt}=k^{+}(R_0-y)Q(x)-k^{-}y.
\end{equation}
Since the integrin dynamics are much faster than the macroscopic time scale phenomena, we assume that they equilibrate rapidly \cite{engwer2016effective,hunt2018dti,conte2021mathematical}. Thus, after rescaling $y / R_0\rightarrow y$, we consider the unique steady state $y^{*}$ of (\ref{reac_eq}), given by

$$y^{*}=\frac{k^{+}Q(x)}{k^{+}Q(x)+k^{-}}=:f(Q(x)),$$
and we define a new internal variable $z := (y-y^{*})\in \mathbf{Z}=(y^{*}-1,y^{*})\subset \mathbb{R}$, which measures the deviation of $y$ from its steady state \cite{engwer2015glioma, hunt2018dti}.\\
Considering the piecewise location of a single cell $x=x_0+vt$ through the density field $Q(x)$, $z$ satisfies

\begin{equation}
	\label{eqn:internal_state}
	\setlength{\jot}{10pt}
	\begin{aligned}
		\frac{dz}{dt}& = -\big(\underbrace{(k^{+}Q(x)+k^{-})z-f^{'}(Q(x))\langle\,v,\nabla_xQ(x)\rangle}_{:=G(t,x,v,z,Q)}\big),
	\end{aligned}
\end{equation}
where $\langle\cdot,\cdot\rangle$ is the scalar product on $\mathbf{V}\times \mathbb{R}^3$ and $f^{'}(Q(x))=\frac{k^{+}k^{-}}{(k^{+}Q(x)+k^{-})^2}$. The internal variable $z$ is bounded as long as $\nabla_x Q(x)$ is bounded and its sign depends on the current orientation of the  cell w.r.t the gradient of $Q(x)$.
\subsubsection{PDifMP description for glioma cell movement}
\label{subsec_PDifMPOnGlioma}
To model cell movement under the influence of external signals, we assume that the sample path of an individual cell starting in position $x_0$ and moving in a certain direction due to contact guidance for a random period of time is given by

\begin{equation}
	\label{SDE_1}
	\left\{
	\begin{array}{ll}
		dx_t & = v_t dt+ \sigma dW_t,\\
		x(0)&=x_0.
	\end{array}
	\right. 
\end{equation}
Here, the second term in the r.h.s represents the stochastic variability in the velocity, with $\sigma \in \mathbb{R}$ being the diffusion coefficient and $W_t$ the standard Wiener process.\\
\indent Due, for instance, to collisions with other cells in their surrounding \cite{loy2020kinetic,painter2002volume}, during the movement a cell stops for a negligible duration and reorients its path \cite{loy2020kinetic}. This causes the cell to adopt a new velocity to continue migrating in the new direction until another obstacle is encountered. To describe this process, we rely on the introduced PDifMP framework. We set $E_1=\mathbf{X}\times \mathbf{Z}\subset \mathbb{R}^{4}$ and we denote by $S_t:=(X_t,Z_t)$ the continuous component describing cell motion. Their evolution is characterised through the SDE (\ref{SDE_1}) for cell motility and the ODE (\ref{eqn:internal_state}) for the interactions with the microenvironment. Both processes are affected by spontaneous velocity changes induced by the jump process $V_t$. Then, we denote by $E=E_1\times \mathbf{V}$ the state space of the piecewise process $U_t=(X_t,Z_t,V_t)$ for cell motility and migration and by ${\phi:[0,T]\times E \rightarrow E_1}$, the solution to the coupled system (\ref{eqn:internal_state})- (\ref{SDE_1}).\\
\indent As the duration of reorientation is negligible, we describe the direction of a cell at a given instant. Moreover, under the additional assumption that the motion is Markovian in the state space, we state that cell direction is described with an inhomogeneous Poisson-like process \cite{gabbiani2017mathematics}, whose intensity depends on time, position on the scaled sphere $\mathbf{V}$, and internal state. Thus, the cell reorientation rate referring to the jump rate function $\lambda:[0,T]\times E \rightarrow \mathbb{R}_{+}$ of the stochastic process $u_t$ depends on the integrin state $z$. This means that the binding process is seen as the onset of reorientation. In particular, following \cite{sun2006regulation}, we assume that, if many integrins are bounded, cells tend to change direction frequently in order to escape the densely packed areas, resulting in an increased rate $\lambda$. Thus, following \cite{engwer2016effective, engwer2016multiscale}, we set $\lambda(u_t):=(\lambda_0-\lambda_1 z_t)\geq 0$\, with $\lambda_0$ and $\lambda_1$ positive constants. In particular, $\lambda_0$ refers to the basal turning frequency of an individual cell \cite{sidani2007} accounting for the "spontaneous" cell motility, while the term $\lambda_1z$ represents the variation of the turning rate in response to environmental signals.\\
\indent Following the construction described in Section \ref{const} with initial state $u_0=(x_0,z_0,v_0)$, we use the jump rate function $\lambda$ defined in (\ref{gen_surv}) to determine the duration of movement before any reorientation of direction occurs. Moreover, considering that the velocity jump process $v_t$ is of Markovian type, we have that cells retain no memory of their velocities before the reorientation. Thus, we define the Markov transition kernel $\mathcal{Q}$, determining the post-velocity jump state of the process $u_t$, using $K(x,v,v^{'})$, which describes the distribution of newly chosen velocities, having that $K(x,v,v^{'})=K(x,v)$.
\begin{definition}
	Let $\nu$ be the standard Lebesgue measure on $(\mathbf{V}, \mathcal{V})$ and $K: \mathbf{X} \times \mathbf{V} \rightarrow [0,\infty]$ be a measurable function with respect to the $\sigma$-algebra $\mathcal{X}\otimes\mathcal{V}$ such that 
	
	\begin{equation}
		\label{poba_density}
		\int_{\mathbf{V}} K(x,v)\nu(dv)=1, \quad \forall x\in \mathbf{X}.
	\end{equation}
	Then, the mapping 
	
	\begin{equation*}
		\left\{
		\begin{array}{ll}
			\mathcal{Q}: \mathbf{X} \times\mathcal{V} \rightarrow [0,1], & ~ \\[0.2cm]
			\mathcal{Q}= \int_{\mathbf{V}} K(x,v)\nu(dv), & ~
		\end{array}
		\right. 
	\end{equation*}
	defines a Markov transition kernel over $\mathbf{V}$, where $\nu(dv)=dv$.
\end{definition}
\noindent Denoting by $q(x,\hat{v})$ the fiber distribution function over $\mathbf{V}$, with $\hat{v}=\frac{v}{\Vert v\Vert}\in \mathbb{S}^{2}$, and by

\begin{equation*}
	w:= \int_{\mathbf{V}} q(x,\hat{v})dv=\alpha^2,
\end{equation*}
a scaling constant \cite{hillen2013transport, hillen20042}, we assume that the dominant directional cue leading cell migration is given by the fiber network. Thus, the transition probability kernel is given by

\begin{equation}
	\label{T_kernel}
	K(x,v)=\frac{q(x,\hat{v})}{w}.
\end{equation}
For the fiber distribution function $q(x,\hat{v})$, different expressions can be found in the literature, such as the Von Mises-Fisher Distribution, the Peanut Distribution Function, or the Orientation Distribution Function (ODF) \cite{painter2013mathematical, aganj2011}. A comparison among these distributions have been proposed in \cite{conte2020glioma}, in both 1D and 2D scenarios. We rely on this analysis and we choose the ODF for describing $q(x,\hat{v})$, i.e., we set

\begin{equation}
	q(x,\hat{v})=\frac{1}{4\pi \mid\mathbb{D}(x)\mid^{\frac{1}{2}}(\hat{v}^T(\mathbb{D}(x))^{-1}\hat{v})^{\frac{3}{2}}}\,.
	\label{q_ODF}
\end{equation}
Here, $\hat{v}$ stands for the fiber direction, $x$ for spatial position within the brain, while $\mathbb{D}$ is the diffusion tensor taking into account information about the water diffusivity in the brain \cite{conte2020glioma}. We also assume that fibers are not polarised, i.e., $q(x,\hat{v})=q(x,-\hat{v})$ for all $\hat{v} \in \mathbb{S}^{2}$.
It is straightforward to verify that $q$ is a probability distribution on $\mathbb{S}^{2}$ \cite{engwer2016effective, engwer2015glioma, engwer2016multiscale}.\\  
\indent From (\ref{gen_surv}), it is possible to construct the sequence of jump times $(T_n)_{n\geq 1}$, with $T_n=\tau_1+\dots+\tau_n$ for all $n\geq 1$ (and $T_0=0$ by convention), such that the process $U_t$ describing cellular movement is piecewise constructed on each interval $[T_i,T_{i+1})$, $i=1,\dots,n$, via the characteristics $(\phi,\lambda,\mathcal{Q})$ given by

\begin{equation}
	\label{charc}
	\left\{
	\begin{array}{ll}
		\phi& = \left( v_t t+ \sigma W_t, z_t \right)^{T},\\[0.2cm]
		\lambda & = \lambda_0-\lambda_1z_t,\\[0.2cm]
		\mathcal{Q} & = \dfrac{1}{w}\displaystyle\int_{\mathbf{V}}q(x,\hat{v})dv.
	\end{array}
	\right. 
\end{equation}
Here, $z_t$ is the solution of (\ref{eqn:internal_state}) and $v_t$ is a piecewise constant over each interval of random length $T_{i+1}-T_i$. As proven in \cite{bujorianu2006toward}, this construction leads to a c\`adl\`ag strong Markov process, describing cell motion in an anisotropic environment.\\
In summary, the overall system describing a contact-mediated movement of glioma cells at the microscopic scale reads

\begin{equation}
	\label{Micro_sys}
	\left\{
	\begin{array}{ll}
		dX_t & = V_tdt+\sigma dW_t,\\[0.2cm]
		dZ_t & =-\big(G(t,X_t,Z_t,V_t,Q)\big)dt,\\[0.2cm]
		dV_t & = 0dt.\\
	\end{array}
	\right. 
\end{equation}
The solution of (\ref{Micro_sys}) is a triple $U_t=(X_t,Z_t,V_t) \in E$, with ${E=E_1\times\mathbf{V}=(\mathbb{R}^3\times \mathbb{R})\times \alpha \mathbb{S}^2}$, and hereafter we will refer to $(X_t,Z_t,V_t)$ as $(x_t,z_t,v_t)$ as we are talking about the sample path of $U_t$.

\subsection{Derivation of the mesoscopic equation and its macroscopic limit}
We rely on the definition of the extended generator of $(U_t)_{t\in[0,T]}$ given in Section \ref{2E} to obtain a mesoscopic equation describing the evolution of the joint probability density function of all microscopic variables.
In the specific, for all test functions $f\in \mathcal{D}(\mathcal{A})$, the extended generator $\mathcal{A}$ of the above defined process $U_t$ reads

\begin{equation}
\label{generator}
   \begin{split}
	&\mathcal{A}f(x,z,v)=\big(\langle\nabla_x f(x,z,v),v_t\rangle-\partial_z(G(t,x,z,v,Q))f(x,z,v)\big)\\[0.1cm] 
	&+\frac{1}{2} \sigma^2 \Tr\left((\nabla_x \nabla^T_x) f(x,z,v)\right)+ \lambda(z) \int_{\mathbf{V}} (f(x,z,v^{'})-f(x,z,v))\mathcal{Q}(x,v,dv^{'}),
   \end{split} 
\end{equation}
where $\lambda$ and $\mathcal{Q}$ are given in (\ref{charc}). Notice that the integral term in (\ref{generator}) is defined over $\mathbf{V}$ as the transition kernel $\mathcal{Q}$ has a density defined on $\mathbf{V}$.\\
\indent Let $g(t,x,z,v)$ be the joint pdf of the microscopic variables at time $t\in [0,T]$, position $x\in \mathbf{X}$, internal state $z\in \mathbf{Z}$, and velocity $v \in \mathbf{V}$. In this context, we refer to as glioma density function. The adjoint operator $\mathcal{A}^{*}g$ is given by

\begin{equation}
    \begin{split}
	\label{Adjoint_generator}
&	\mathcal{A}^{*}g(x,z,v)=-\big(\langle\nabla_x g(x,z,v),v_t\rangle-\partial_z(G(t,x,z,v,Q))g(x,z,v)\big)\\[0.1cm]
&+\frac{1}{2} \sigma^2 \Tr\left((\nabla_x \nabla^T_x) g(x,z,v)\right)+ \lambda(z) \int_{\mathbf{V}} (g(x,z,v^{'})-g(x,z,v))\mathcal{Q}(x,v,dv^{'}).
	\end{split}
\end{equation}

Thus, following the analysis of Section \ref{GFPE}, the generalised Fokker-Planck equation for the evolution of $g(t,x,z,v)$ reads

\begin{equation}
	\label{meso_comp}
	\begin{split}
		&\partial_t g(t,x,z,v)+ \langle \nabla_xg(t,x,z,v),v\rangle-\partial_z(G(t,x,z,v,Q))g(t,x,z,v)\\[0.1cm]
		&-\frac{1}{2} \sigma^2 \Tr\left((\nabla_x \nabla^T_x) g(t,x,z,v)\right) =\mathcal{L}g(t,x,z,v),
	\end{split}
\end{equation}
where, from (\ref{T_kernel}), the turning operator reads \cite{othmer2000diffusion, hillen20052, painter2013mathematical, engwer2016effective}

\begin{equation}
	\label{Turning_Operator}
	\mathcal{L}g(x,z,v)= \lambda(z) \int_{\mathbf{V}} (g(x,z,v^{'})-g(x,z,v))\frac{q(x,v)}{w}dv^{'},
\end{equation}
\begin{remark}
	Note that for $\sigma=0$, (\ref{meso_comp}) coincides with the kinetic transport equation derived in \cite{engwer2016effective, hunt2018dti, engwer2015glioma}. This means that the PDMP resulting from setting $\sigma=0$ in (\ref{meso_comp}) is the formally defined mathematical model underlying the description in \cite{engwer2016effective, hunt2018dti, engwer2015glioma}. 
\end{remark}
\noindent We introduce the notations

\begin{align*}
	\mathbb{E}_q(x):=&\int_{\mathbb{S}^{2}}\hat{v} q(x,\hat{v})d\hat{v},\\
	\mathbb{V}_q(x):=&\int_{\mathbb{S}^{2}}(\hat{v} -\mathbb{E}_q)\otimes (\hat{v} -\mathbb{E}_q) q(x,\hat{v})d\hat{v},
\end{align*}
for the mean fiber orientation and the variance-covariance matrix of the fiber orientation distribution, respectively. Notice that the symmetry on the fiber distribution implies $\mathbb{E}_q=0$.\\
\indent Following \cite{engwer2016effective, hunt2018dti}, we model proliferation as an effect of cell-tissue interactions via integrin binding

\begin{equation}
	\label{proliferation}
	\mathcal{P}(g(t,x,z,v))=\mu(M(t,x))\int_{\mathbf{Z}} \mathcal{X}(x,z,z^{'})g(t,x,v,z^{'})Q(x)dz^{'}.
\end{equation}
Here, $M(t,x)$ denotes the macroscopic cell density, that is, the marginal distribution of $g(t,x,z,v)$ over all possible velocities and internal states, i.e.,

$$M(t,x)=\int_{\mathbf{V}}\int_{\mathbf{Z}}g(t,x,z,v)dzdv.$$
Moreover, $\mu(M)$ is the growth function and the kernel $\mathcal{X}(x,z,z^{'})$ is a probability density in the second variable $z$ characterising the transition from $z^{'}$ to $z$ during the proliferation process at position $x$. For $\mathcal{X}$ we only assume that the operator $\mathcal{P}(g)$ is uniformly bounded in the $L^2$-norm, a reasonable biological condition related to the space-imposed limits on cell division.
Thus, for the evolution of $g(t,x,z,v)$ we obtain the following equation

\begin{equation}
   \label{mesoscopic_eq}
       \begin{split}
		&\partial_tg(t,x,z,v)+\langle \nabla_x g(t,x,z,v), v \rangle-\partial_z(G(t,z,Q)g(t,x,z,v))\\[0.1cm]
		&-\frac{1}{2} \sigma^2 \Tr\left((\nabla_x \nabla^T_x)g(t,x,z,v)\right) 	=\mathcal{L}g(t,x,z,v)+\mathcal{P}(g(t,x,z,v)).
		\end{split}
\end{equation}
\noindent Due to the high dimensionality of (\ref{mesoscopic_eq}), numerical simulations of this equation would be too expensive. Moreover, clinicians are more interested in the macroscopic dynamics of the tumour rather than in the lower scale interactions. Thus, we derive the macroscopic equation for the evolution of the tumour density, based on the definition of the moments of $g$ with respect to $v$ and $z$:

\begin{align*}
	m(t,x,v)&=\int_{\mathbf{Z}} g(t,x,z,v)dz           &  M(t,x)&=\int_{\mathbf{V}} m(t,x,v)dv              \\[0.1cm]
	m^z(t,x,v)&=\int_{\mathbf{Z}} zg(t,x,z,v)dz         &  M^z(t,x)&=\int_{\mathbf{V}} m^z(t,x,v)dv.
\end{align*}
Notice that we do not consider higher order moments of $g$ with respect to $z$ as the subcellular dynamics are much faster than the events taking place on the other scales, so that the deviation $z$ is close to zero. Dropping the $(t,x)$ notation for simplicity, the moment equations reads

\begin{equation}
\label{eqn:Em}
\begin{split}
	&\partial_t m+\langle \nabla_x m, v\rangle- \frac{1}{2} \sigma^2 \Tr\left((\nabla_x \nabla^T_x) m\right)=  -\lambda_0 m + \lambda_1 m^z + \lambda_0 \frac{q}{w}M -
	\lambda_1 \frac{q}{w}M^z \\[0.1cm]
	&+\mu(M)\int_{\mathbf{Z}} \int_{\mathbf{Z}} \mathcal{X}(x,z,z^{'})g(t,x,v,z^{'})Q(x)dz^{'}dz,
\end{split}
\end{equation}
and 

\begin{equation}
	\label{eqn:Em_z}
	\begin{split}
	\hspace{-0.2cm}&\partial_t m^z+\langle\nabla_x m^z,v\rangle- \frac{1}{2}\sigma^2 \Tr\left((\nabla_x \nabla^T_x) m^z\right)= -(Q(x)k^{+}+k^{-})m^z 
	+ \lambda_0 \frac{q}{w}M^z\\[0.1cm]
	&-\lambda_0 m^z +f^{'}(Q(x))\langle v,\nabla_x Q(x)\rangle m  + \mu(M)\int_{\mathbf{Z}} \int_{\mathbf{Z}} z \mathcal{X}(x,z,z^{'})g(t,x,v,z^{'})Q(x)dz^{'}dz.
	\end{split}
\end{equation}
Following \cite{engwer2015glioma,engwer2016effective}, we consider a parabolic scaling of the moment equations setting $ x \mapsto \epsilon x$ and  $t \mapsto \epsilon ^2 t$ for space and time variables, respectively. In particular, we scale the growth rate function $\mu(M)$ with $\epsilon^2$ as it accounts for faster dynamics. Thus, we obtain

\begin{equation}
	\label{eqn:Eps}
	\begin{split}
		&\epsilon^2 \partial_t m+\epsilon \langle \nabla_x m, v \rangle- \epsilon^2\frac{1}{2}\sigma^2 \Tr\left((\nabla_x \nabla^T_x) m\right)=-\lambda_0 m + \lambda_1 m^z +\lambda_0 \frac{q}{w}M \\[0.2cm]
		& 
		-\lambda_1 \frac{q}{w}M^z + \epsilon^2\mu(M)\int_{\mathbf{Z}} \int_{\mathbf{Z}} \mathcal{X}(x,z,z^{'})g(t,x,v,z^{'})Q(x)dz^{'}dz.
	\end{split}
\end{equation}
and 

\begin{equation}
	\label{eqn:Eps_z}
	\begin{split}
		&\epsilon^2 \partial_t m^z+\epsilon \langle\nabla_x m^z, v \rangle- \epsilon^2\frac{1}{2} \sigma^2 \Tr\left((\nabla_x \nabla^T_x) m^z\right)=\\[0.2cm]
		&- (Q(x)k^{+}+k^{-})m^z + \epsilon f^{'}(Q(x)) \langle v,\nabla_x Q(x)\rangle m  -\lambda_0 m^z  + \lambda_0 \frac{q}{w}M^z  \\[0.2cm]
		&+\epsilon^2\mu(M)\int_{\mathbf{Z}} \int_{\mathbf{Z}} z \mathcal{X}(x,z,z^{'})g(t,x,v,z^{'})Q(x)dz^{'}dz.
	\end{split}
\end{equation}
We consider the \textit{Hilbert expansion methods} \cite{ellis1973chapman, hunt2018dti} expanding the moments of $g$ as

\begin{align*}
	m(t,x,v)&=\sum_{k=0}^{\infty} \epsilon^{k}m_k\,,       &  M(t,x)&=\sum_{k=0}^{\infty}  \epsilon^{k}M_k \,,            \\
	m^z(t,x,v)&=\sum_{k=0}^{\infty} \epsilon^{k}m^z_k\,,           &  M^z(t,x)&=\sum_{k=0}^{\infty} \epsilon^{k}M^z_k. 
\end{align*}
By equating the same powers of $\epsilon$ in  (\ref{eqn:Eps}) and (\ref{eqn:Eps_z}), we derive the equation for the leading order coefficient $M_0$ of the Hilbert expansion of $M$. Thus we obtain \\[0.1cm]

\noindent $\epsilon^0$:
\begin{align}
	0= & -\lambda_0m_0+\lambda_1m^z_0
	+\lambda_0\frac{q}{w}M_0 -\lambda_1 \frac{q}{w}M^z_0         \\[0.2cm]
	0= &-(Q(x)k^{+}+k^{-})m^z_0-\lambda_0m^z_0+\lambda_0\frac{q}{w}M^z_0.
\end{align}

\noindent $\epsilon^1$:
\begin{align}
	\label{epsil_11}
	\langle\nabla_x m_0,v\rangle= & -\lambda_0m_1+\lambda_1m^z_1+\lambda_0\frac{q}{w}M_1-\lambda_1 \frac{q}{w}M^z_1      \\[0.2cm]
	\langle\nabla_x m^z_0,v\rangle= &  -(Q(x)k^{+}+k^{-})m^z_1+f^{'}(Q(x))\langle v,\nabla_x Q(x)\rangle m_0 -\lambda_0m^z_1+\lambda_0\frac{q}{w}M^z_1.
	\label{epsil_12}
\end{align}

\noindent $\epsilon^2$:
\begin{equation}
	\label{epsilon_2}
	\begin{split}
&	\partial_tm_0+\langle\nabla_x m_1,v\rangle-\frac{1}{2} \sigma^2 \Tr\left((\nabla_x \nabla^T_x) m_0\right)= -\lambda_0m_2+\lambda_1m^z_2+\lambda_0\frac{q}{w}M_2\\[0.2cm]
& -\lambda_1 \frac{q}{w}M^z_2+\mu(M_0)\int_{\mathbf{Z}}\int_{\mathbf{Z}}\mathcal{X}(x,z,z^{'})g(t,x,v,z^{'})Q(x)dz^{'}dz.
	\end{split}
\end{equation}
With classical scaling arguments (see \cite{engwer2015glioma} for more details), we obtain ${M^z_0=m^z_0=0}$ and $m_0=\frac{q(x,v)}{w}M_0$. On account of that, using the symmetry assumption, i.e., $\mathbb{E}_q=0$, from (\ref{epsil_12}), we obtain ${M^z_1=0}$, and 

\begin{equation*}
	m^z_1=\frac{f^{'}(Q(x))\langle v,\nabla_xQ(x)\rangle\frac{q}{w}M_0}{\lambda_0+Q(x)k^{+}+k^{-}}.
\end{equation*}
Moreover, considering (\ref{epsil_11}) and following the analysis in \cite{othmer2000diffusion,engwer2015glioma}, we get ${M_1=0}$, and 

$$m_1=\frac{1}{\lambda_0} \Big[- \langle\nabla_xm_0,v\rangle+ \lambda_1 \Big( \frac{f^{'}(Q(x))\langle v,\nabla_x Q(x)\rangle\frac{q}{w}M_0}{\lambda_0+Q(x)k^{+}+k^-} \Big)  \Big].$$
Replacing it into (\ref{epsilon_2}) and integrating over $\mathbf{V}$, we get:

\begin{equation}
	\partial_tM_0+\int_{\mathbf{V}}\langle\nabla_x m_1,v\rangle dv-\frac{1}{2}\int_{\mathbf{V}} \sigma^2 \Tr(\nabla_x \nabla^T_x) (m_0)dv=\mu(M_0)Q(x)M_0,
\end{equation}
where

\begin{equation*}
       \frac{1}{2}\sigma^2\int_{\mathbf{V}}  \Tr(\nabla_x \nabla^T_x) (m_0)dv= \frac{1}{2}\sigma^2\int_{\mathbf{V}}  \Tr(\nabla_x \nabla^T_x) \big( \frac{q(x,v)}{w}M_0 \big) dv=\frac{1}{2}\sigma^2 \Delta (M_0). 
\end{equation*}

\noindent Therefore, the evolution equation for $M_0$ reads

	\begin{equation*}
	    \begin{split}
	      &\partial_tM_0 -\nabla_x\cdot\big(D_T(x)\nabla_xM_0\big)+\nabla_x\cdot\big(D_T(x)l(Q(x))\nabla_xQ(x)M_0-P_T(x)M_0\big)\\[0.1cm]
		&-\frac{1}{2}\sigma^2 \Delta (M_0)=\mu(M_0)Q(x)M_0,
	    \end{split}
	\end{equation*}
where

\begin{equation}
	l(Q(x)):= \frac{\lambda_1f^{'}(Q(x))}{\lambda_0+Q(x)k^{+}+k^-},
\end{equation}
denotes the function that carries the information about the influence of the subcellular dynamics, while

\begin{equation}
	D_T(x):=\frac{1}{\lambda_0} \int_{\mathbf{V}} \frac{q(x,v)}{w}v\otimes vdv,
	\label{DT_def}
\end{equation}
refers to the macroscopic tumour diffusion tensor. In addition, the tumour drift velocity is given by

\begin{equation}
	P_T(x):=\frac{1}{\lambda_0}\int_{\mathbf{V}} \nabla_x\left(\frac{q(x,v)}{w}\right) v\otimes vdv.
\end{equation}
In view of the results obtained in \cite{engwer2015glioma}, the $\epsilon$-correction terms for $M$ can be left out and, after ignoring the higher order terms and discarding subscripts, we obtain the following evolution equation characterising the macroscopic glioma density:

\begin{equation}
	\begin{split}
		&\partial_tM -\nabla_x.\big(D_T(x)\nabla_xM\big)+\nabla_x.\big(D_T(x)l(Q(x))\nabla_xQ(x)M-P_T(x)M\big)\\[0.2cm]
		&-\frac{1}{2}\sigma^2 \Delta M=\mu(M)Q(x)M.
	\end{split}
	\label{macro_M_eq}
\end{equation}

\noindent Using the theory of monotone operators for nonlinear parabolic equations and following the approach in \cite{showalter2013monotone,ruuvzivcka2004fixpunktsatze}, it is possible to prove the existence, uniqueness and non-negativity of the solution of the following parabolic problem with homogeneous Neumann boundary conditions.

\small\begin{align}
\hspace{-0.5cm}	\label{Well_poss}
 \begin{cases*}
	\partial_tM\!\! -\nabla_x\!\!\cdot\!\left(\!\left(\!D_T(x)\!\!+\!\!\dfrac{1}{2}\sigma^2\right)\nabla_xM\!+\!\gamma(D_T(x),Q(x))M\!\!\right)\!\!-\Gamma(M,Q(x))=0, & in $[0,T]\times  \mathbf{\Omega}$,\\[0.2cm]
		\nabla_xM\cdot \hat{n} = 0  & on $[0,T]\times \partial \mathbf{\Omega}$,\\[0.2cm]
		M(0,x)= \Tilde{M}_0(x),   & in $\mathbf{\Omega}$,
	\end{cases*}
\end{align}
\normalsize
where

\begin{equation*}
	\begin{split}
		&\gamma(D_T(x),Q(x)):=P_T(x)-D_T(x)l(Q(x))\nabla_xQ(x),\\[0.2cm]
		&\Gamma(M,Q(x)):=\mu(x,M)Q(x)M.
	\end{split}
\end{equation*}
We refer the reader to Appendix \ref{Appendix_A} for more details about the necessary assumptions on the operators and for an outline of the proof of the well-posedness of the macroscopic problem (\ref{Well_poss}).

\section{Numerical simulations}
\label{section4}
We perform 2D simulations of the macroscopic equation for the tumour cells \eqref{macro_M_eq} to study the impact of both the subcellular dynamics and the stochastic parameter $\sigma$ on the overall tumour evolution.\\
\indent With this aim, we firstly specify parameters and coefficient functions involved in the equation. Concerning the tumour diffusion tensor $D_T(x)$ in \eqref{DT_def}, we numerically compute it using the orientation distribution function given in \eqref{q_ODF}, 
where $\mathbb{D}(x)$ represents the water diffusion tensor obtained from processing (patient-specific) DTI data. Taking advantage of this DTI information, for the macroscopic tissue density $Q(x)$ we assume the following expression

\begin{equation}
	Q(x)=FA(\mathbb{D}(x))\,,
	\label{Q_def}
\end{equation}
where $FA$ refers to the fractional anisotropy of the tissue. We refer to \cite{engwer2015glioma} for its definition. This choice is motivated by the fact that the fractional anisotropy represents a measure of the fiber alignment and, since in this setting fiber alignment is guiding cell migration, it is reasonable to assume that the function $Q(x)$ expresses higher values where the tissue is more anisotropic.\\
\indent Following several previous works (e.g. see  \cite{engwer2016effective,conte2020glioma}), for the growth rate $\mu(M)$ we employ a logistic growth term defined as

\[
\mu_M=\mu_0\left(1-\dfrac{M}{K_M}\right),
\]
with $\mu_0$ the constant growth coefficient and $K_M$ the tumour carrying capacity. Finally, we report in Table \ref{parameter_mod1} the range for the constant parameter values involved in the macroscopic setting \eqref{macro_M_eq}. The values for the stochastic parameter $\sigma$ are proposed based on the ranges of the other parameters. \\ 

\begin{figure}[h!]
\centering
\includegraphics[width=0.9\textwidth]{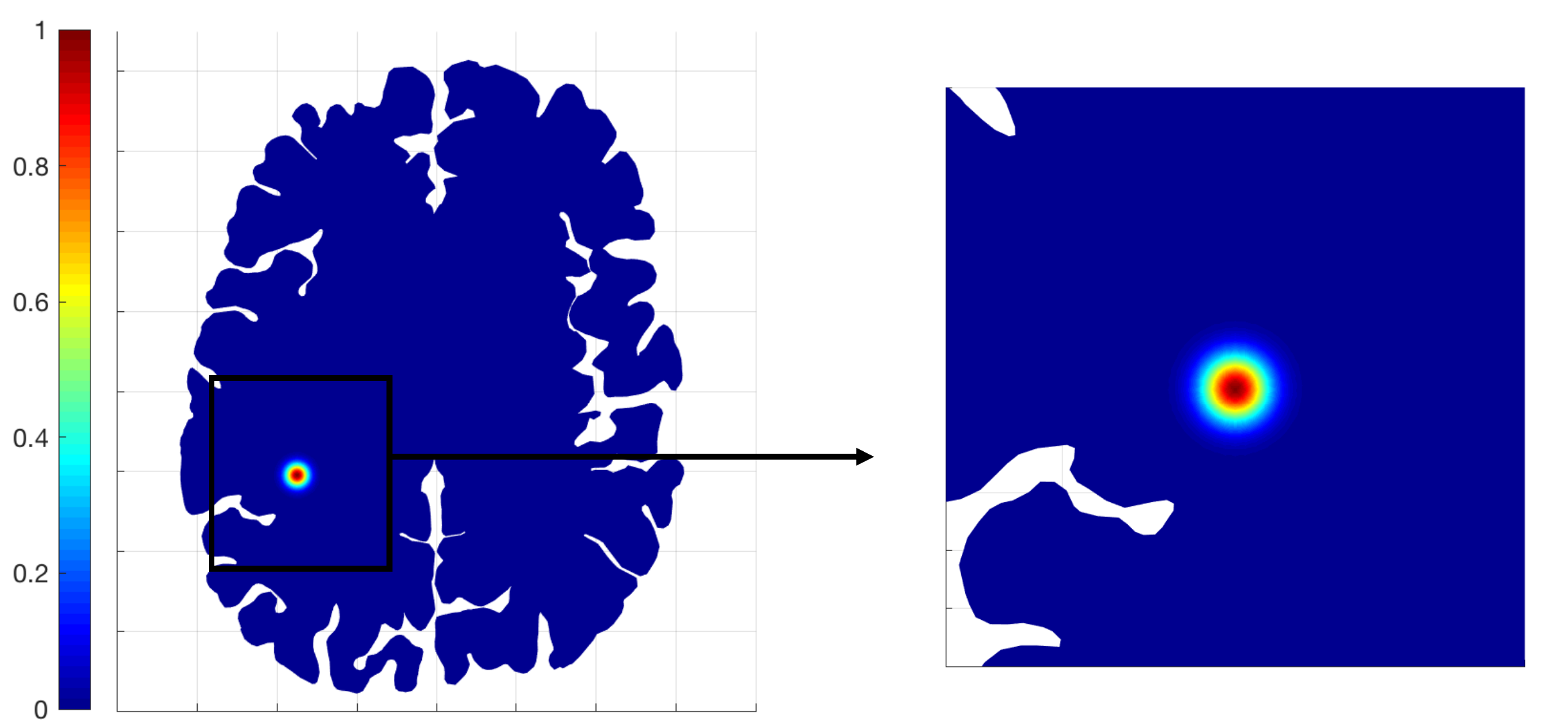}
\caption{{\bf Initial condition of equation \eqref{macro_M_eq}.}}
\label{In_Con}
\end{figure}

\begin{table} [H]
	\begin{center}
		\begin{tabular}{c|c|c|c} 
			\hline  
			\rule{0pt}{3ex}Parameter & Description & Value (unit) & Source \\[1ex]
			\hline
			\rule{0pt}{2ex} $s$ & speed of tumour cells    & $0.21 \cdot 10^{-3}$ (mm$\cdot$ s$^{-1}$) &\cite{chicoine1995}\\[1.5ex]
			\rule{0pt}{2ex} $\lambda_0$  & turning frequency in $\mathcal{L}$ &$[0.25,5]$ (s$^{-1}$)& based on \cite{sidani2007}\\[1.5ex] 
			\rule{0pt}{2ex} $\lambda_1$  & turning frequency in $\mathcal{L}$ &$[-5,5]$ (s$^{-1}$)& based on \cite{engwer2016effective}\\[1.5ex]
			\rule{0pt}{2ex} $\mu_0$ &tumour proliferation rate & $8.44\cdot 10^{-7}$ (s$^{-1}$) &\cite{hunt2018dti}\\[1.5ex]
			\rule{0pt}{2ex} $K_M$ & tumour carrying capacity  &$\approx 10^{6}$ (cells$\cdot$ mm$^{-3}$) & \cite{TCs_dim}\\[1.5ex]
			\rule{0pt}{2ex} $\sigma$ & free stochastic parameter  & $[0.01-0.2]$ (mm$^2\cdot$ s$^{-1}$) & proposed range \\[1.5ex]
			\hline
		\end{tabular}
\end{center}
\vspace{0.3cm}
 \caption{\footnotesize{{\textbf{Model parameters}}}.}
\label{parameter_mod1}
\end{table}
\vspace{-0.5cm}
\indent We present 2D numerical simulations performed with a self-developed code in Matlab (MathWorks Inc., Natick, MA). The computational domain is a horizontal brain slice reconstructed from MRI scans. The DTI dataset used to compute the $D_T(x)$ was acquired at the Hospital Galdakao-Usansolo (Galdakao, Spain), and approved by its Ethics Committee: all the methods employed were in accordance to approved guidelines. A Galerkin finite element scheme for the spatial discretisation is considered, together with an implicit Euler scheme for the time discretisation. For the initial condition, we consider a Gaussian-like aggregate of tumour cells centered at 
$(x_0, y_0) = (-35, -41)$, situated in the left-bottom part of the brain slice. To be specific,

\[
M_0=e^{\frac{(x-x_0)^2+(y-y_0)^2}{8}}\, .
\]
Figure \ref{In_Con} shows the initial condition on the entire 2D brain slice and in the corresponding zoomed region $\mathbf{\Omega}=[-60, -10]\times [-65, -15]$. 

\begin{figure}[h!]
\centering
\includegraphics[width=0.65\textwidth]{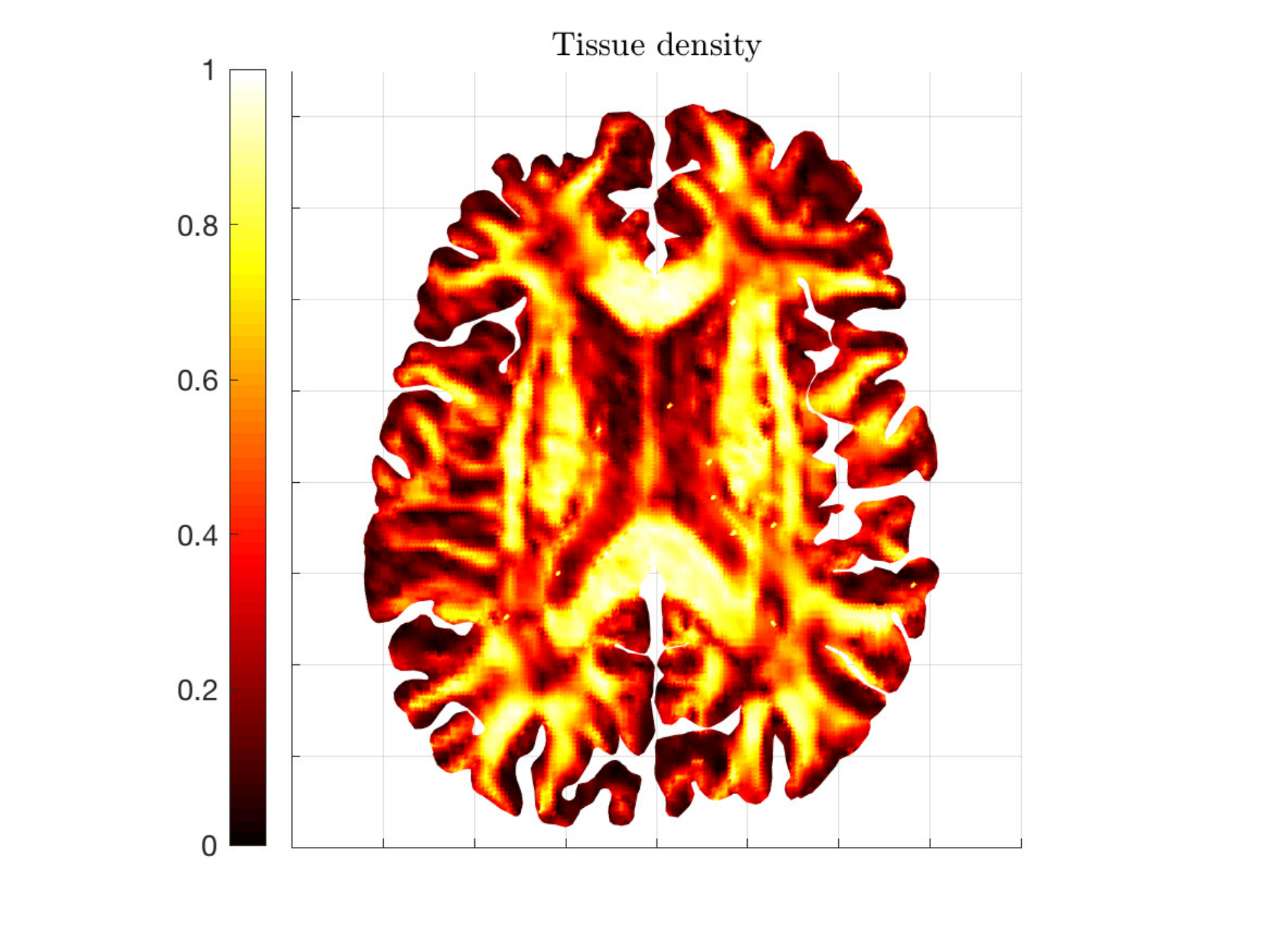}
\caption{{\bf Healthy tissue density}.}
\label{Q_in}
\end{figure}

\noindent Moreover, Figure \ref{Q_in} shows the initial tissue density estimated with \eqref{Q_def}. In particular, yellow areas refer to regions where the fibers are highly aligned and, thus, the value of $FA(\mathbb{D}(x))$ is closer to one, while black-red areas refer to more isotropic regions, where the fibers are randomly distributed \cite{engwer2015glioma}.

\noindent We present different sets of simulations to obtain insight into several features characterising the proposed approach. In detail,
\begin{itemize}
\item[{\bf(A)}] we consider the model for $\sigma=0$ and we evaluate the effects of the variation of $\lambda_1$ and $\lambda_0$ on tumour evolution;
\item[{\bf(B)}] we fix the value of $\lambda_0$ and $\lambda_1$ and we assess the effects of the variation of $\sigma$ on tumour evolution, i.e., the role of the stochastic parameter in the overall dynamics;
\item[{\bf(C)}] we consider different combinations of $\lambda_1$ and $\sigma$ and we show how their respectively effects merge;
\item[{\bf(D)}] following the approach proposed in \cite{bogdanska2017}, we discuss the effects of $\lambda_0$, $\lambda_1$, and $\sigma$ on the estimation of the onsets of malignant transformation from low grade to high grade gliomas.
\end{itemize}

\begin{figure}[h!]
\centering
\includegraphics[width=0.9\textwidth]{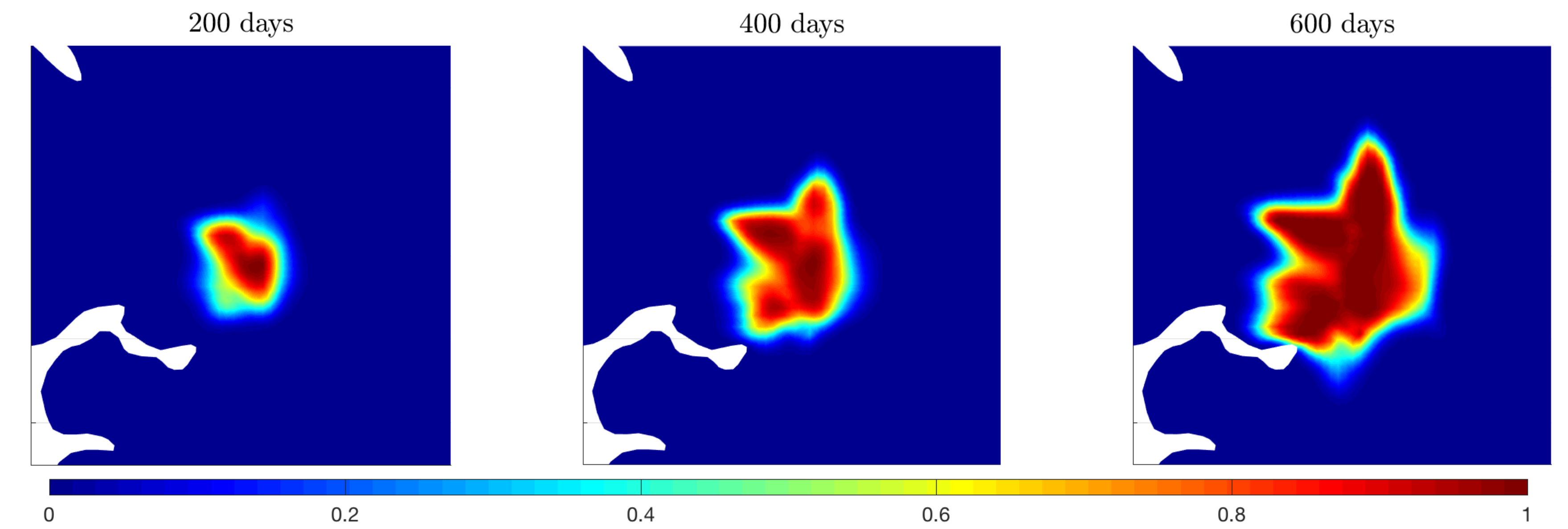}
\caption{{\bf System evolution.} Numerical simulation of equation (\ref{macro_M_eq}) with the parameters listed in Table \ref{parameter_mod1} and for $\lambda_0=\lambda_1=0.8$ (s$^{-1})$. The tumour evolution is shown after 200, 400, and 600 days.}
\label{Evo_lambda1_08}
\end{figure}

\indent Starting from the numerical test {\bf(A)}, we analyse the effects of varying $\lambda_1$ (referring to it as experiment {\bf A.1}) and $\lambda_0$ (referring to it as experiment {\bf A.2}). These experiments are motivated by the fact that obtaining a clear biological estimation for $\lambda_0$ and, especially, for $\lambda_1$ is quite difficult. Thus, understanding the impact of their variation becomes a fundamental point to address. As described in Section \ref{subsec_PDifMPOnGlioma}, $\lambda_0$ refers to the basal turning frequency of an individual cell, while $\lambda_1$ takes into account the role of the receptor dynamics in the evolution. Recalling the expression of the turning rate $\lambda(z)$, we could describe the constant parameters $\lambda_0$ and $\lambda_1$ as the weights of the receptors-independent and receptors-dependent cell turning, respectively. Starting from the analysis on the parameter $\lambda_1$ and in line with some studies concerning the effects of its variability \cite{hunt2018dti} on tumour evolution, we consider the range $\lambda_1\in[-5,5]$ (s$^{-1}$) and we assess the effects of changes in both its sign and modulus. Considering that the turning rate $\lambda(z)=\lambda_0-\lambda_1 z$ has to be non-negative, we should ensure that $\lambda_0\ge\lambda_1 z$, meaning that
\begin{itemize}
\item if $\lambda_1 \ge 0$, the non-negativity is ensured for $\lambda_0\ge \lambda_1/2$;
\item if $\lambda_1 \le 0$, the non-negativity is ensured for $\lambda_0\ge \lambda_1$.
\end{itemize}
Thus, to obtain reasonable values of the turning rate, we should assume $\lambda_1\le\lambda_0$. Although we are aware that negative values of these parameters are not sustained by biological observations, we also include them in our analysis because we want to assess the sensitivity of our results to these parameter changes. In Figure \ref{Evo_lambda1_08}, we firstly show the evolution of the tumour density over time in the limit case in which $\lambda_0=\lambda_1=0.8 \,(\text{s}^{-1})$. 

\begin{figure}[h!]
\centering
\includegraphics[width=0.9\textwidth]{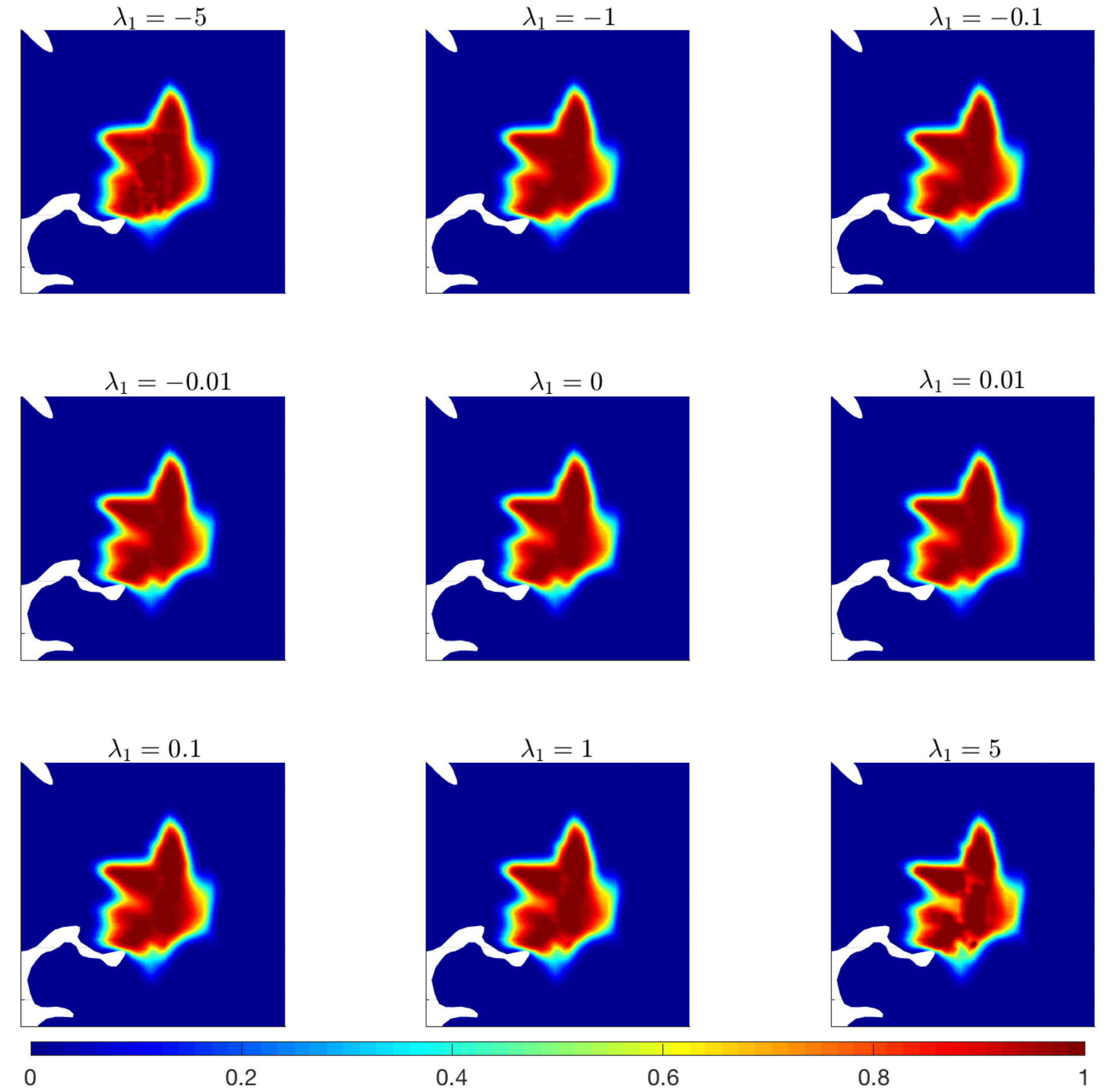}
\caption{{\bf Experiment (A.1).} Numerical simulations of equation (\ref{macro_M_eq}) with the parameters listed in Table \ref{parameter_mod1}, $\lambda_0=0.8 \,(\text{s}^{-1})$ and for different values of $\lambda_1\,(\text{s}^{-1})$. The tumour evolution is shown after 600 days.}
\label{Evo_vary_lambda1}
\end{figure}

\noindent We notice how cells spreading is highly influenced by the underlying fiber structure. Cells clearly tend to move along preferential directions, determined by the fiber bundles, and this gives rise to a heterogeneous tumour mass with an irregular shape, which is a common characteristic for this kind of brain tumours.\\ 
\indent Referring to the tumour situation at the last time step, i.e., after 600 days, we compare the tumour evolution for different values of the parameter ${\lambda_1 \in [-5, 5]}$ (s$^{-1}$), as described in experiment {\bf A.1}. Results are shown in Figure \ref{Evo_vary_lambda1}.

\begin{figure}[h!]
\centering
\includegraphics[width=0.9\textwidth]{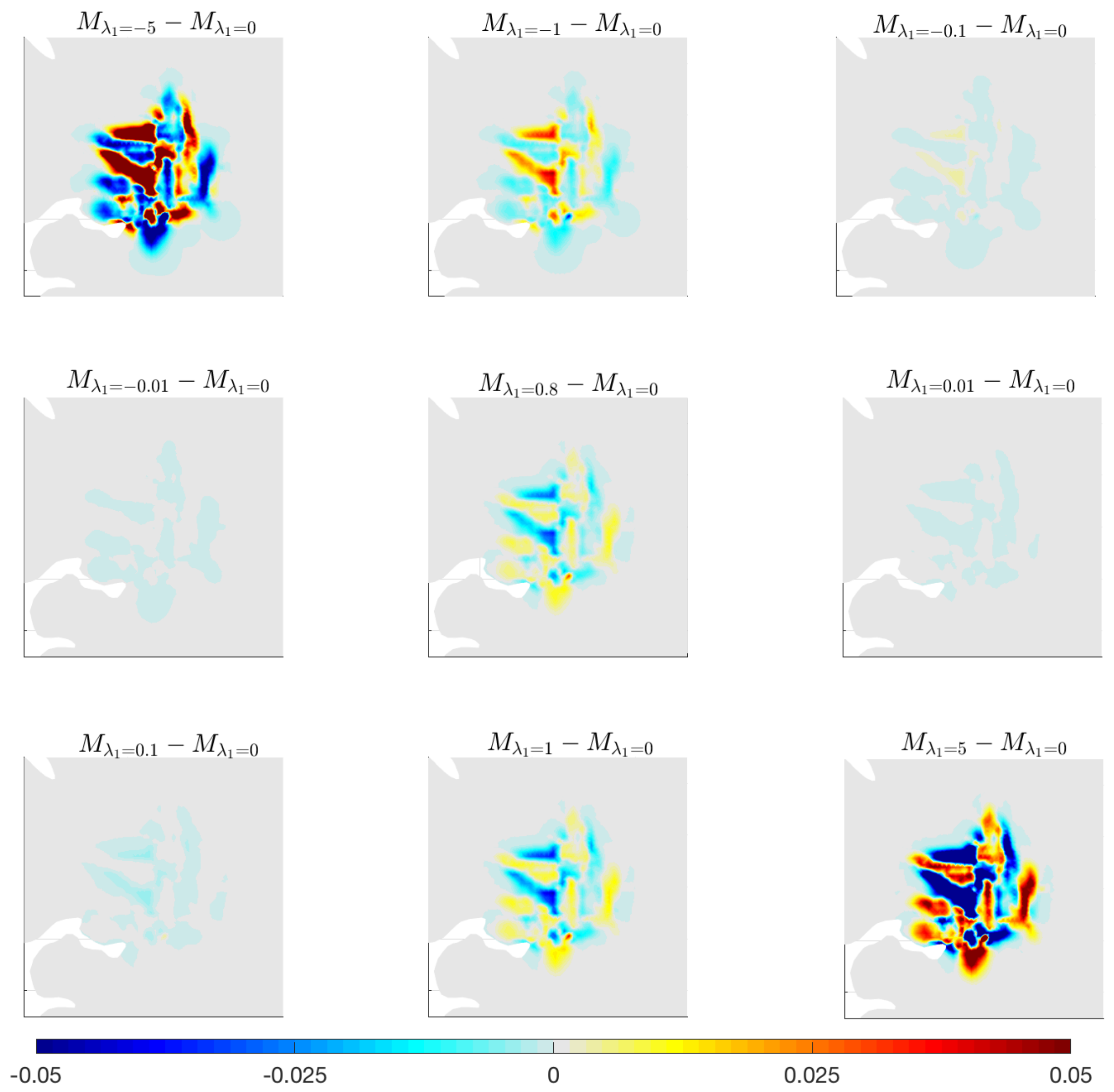}
\caption{{\bf Details of experiment (A.1).} Differences between the solution of system \eqref{macro_M_eq} with $\lambda_1=0$ and the solution obtained for $\lambda_1$ varying in the interval $[-5,5]$ (s$^{-1})$. Results are shown after 600 days. Here $\lambda_0=0.8\, (\text{s}^{-1})$, while the remaining parameters are taken from Table \ref{parameter_mod1}.}
\label{Diff_vary_lambda1}
\end{figure}

\noindent The main effect of varying $\lambda_1$ consists in obtaining a greater or lower level of heterogeneity in the distribution of the tumour cells inside the tumour mass. The external border of the neoplasia, in fact, does not seem to be particularly affected, while the internal dissemination of the cells shows evident changes when $\lambda_1$ varies from large-negative values to large-positive values. In particular, clear differences with respect to the case $\lambda_1=0$ can be observed for quite large values of the parameter ($\vert \lambda_1\vert>1$), while the evolution is qualitatively similar in the cases $\vert\lambda_1\vert<1$. Such differences can be better observed in Figure \ref{Diff_vary_lambda1}, where the differences between the solution of system \eqref{macro_M_eq} for ${\lambda_1=0 \, (\text{s}^{-1}})$ and the solution of the same system for the different values of $\lambda_1$ used in Figure \ref{Evo_vary_lambda1} are shown.

\begin{figure}[h!]
\centering
\includegraphics[width=0.85\textwidth]{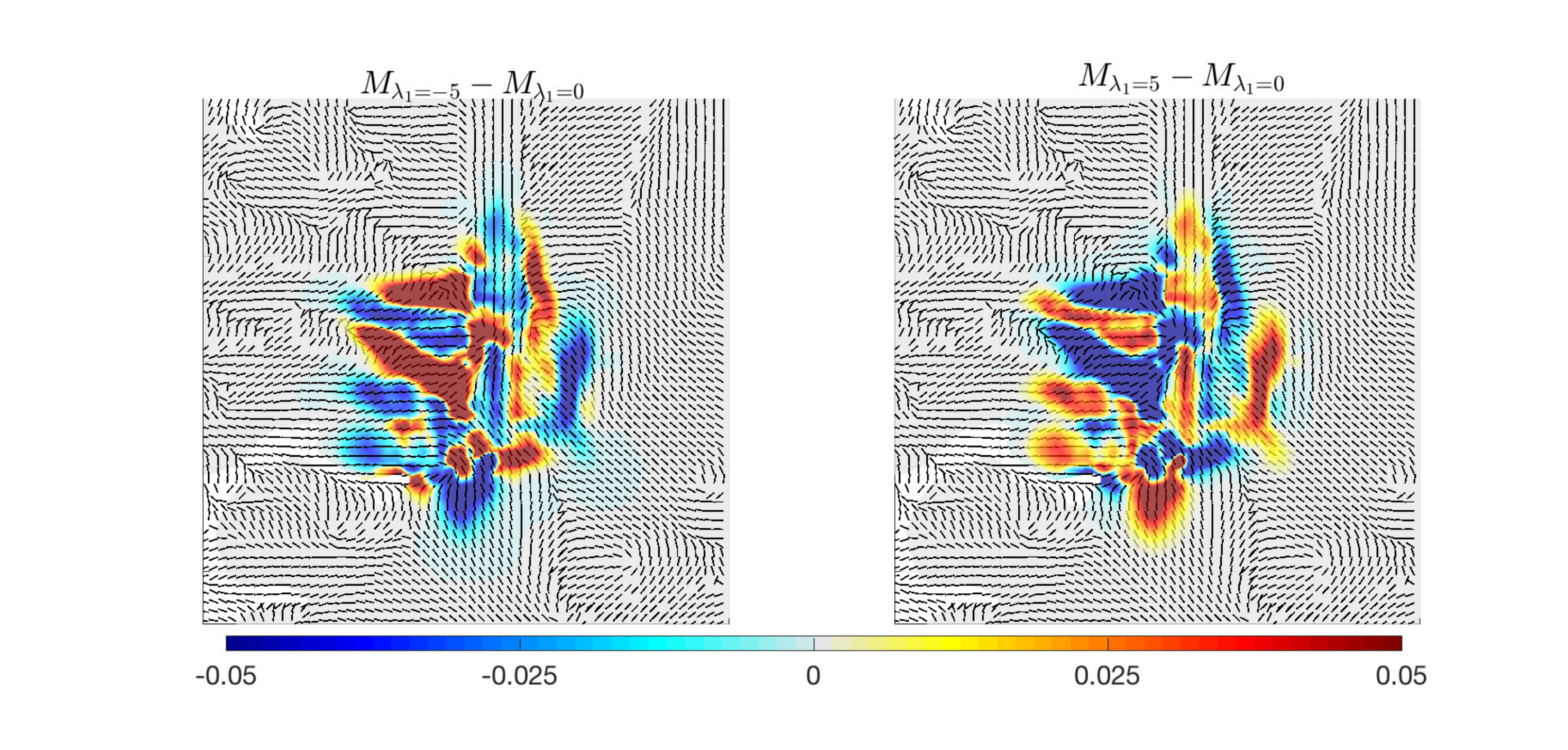}
\caption{{\bf Details of experiment (A.1).} Differences between the solution of system \eqref{macro_M_eq} with $\lambda_1=0$ and the one obtained for $\lambda_1=-5$ \,(\text{s}$^{-1})$ (left plot) and $\lambda_1=5$ \,(\text{s}$^{-1})$ (right plot) after 600 days. The differences are plotted against the fiber direction.}
\label{Fiber_diff_lambda1}
\end{figure}

\noindent The impact of $\lambda_1$ variation can be immediately grasped. There is a clear difference in the spreading inside the tumour mass and in the cell response to the anisotropy of the brain tissue. The impact becomes stronger when $\lambda_1$ increases in modulus, and especially for $\vert\lambda_1\vert>1$. In this case, in fact, the haptotactic component of the dynamics is stronger (in an attractant or repellent way, depending on the sign of $\lambda_1$) and, thus, the heterogeneity of the underlying brain tissue have a larger impact on the dynamics. The mechanism that drives cell migration along the tissue structure can be visualised in details in Figure \ref{Fiber_diff_lambda1}, where the leading eigenvector of the tensor $D_T(x)$ (related to the fiber direction) is plotted together with the differences in the tumour density at 600 days for ${\lambda_1=5 \,(\text{s}^{-1}})$ and ${\lambda_1=-5 \,(\text{s}^{-1}})$.

\begin{figure}[h!]
\centering
\includegraphics[width=0.7\textwidth]{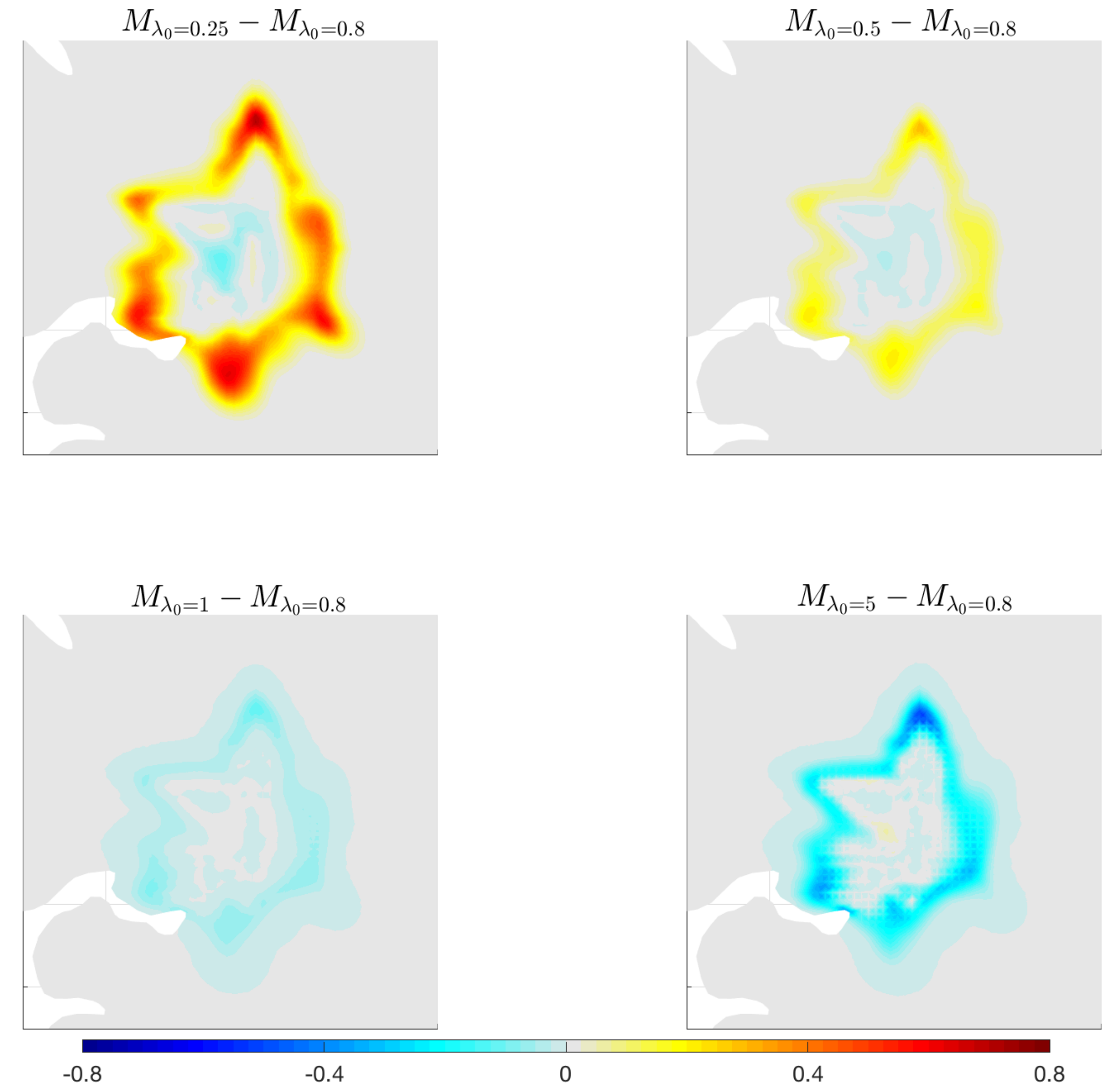}
\caption{{\bf Experiment (A.2).} Differences between the solution of system \eqref{macro_M_eq} with ${\lambda_0=0.8 \,(\text{s}^{-1})}$ and the one obtained for $\lambda_0$ varying in the interval $[0.25,5]$ (\text{s}$^{-1})$. Results are shown after 600 days. The remaining parameters are taken from Table \ref{parameter_mod1}, while ${\lambda_1=0.8\,(\text{s}^{-1})}$.}
\label{Diff_vary_lambda0}
\end{figure}

\noindent Recalling the expression given for the tissue density \eqref{Q_def}, from the left plot of Figure \ref{Fiber_diff_lambda1} we notice that, where the fibers are strongly aligned (e.g. along the central vertical bound), we obtain negative values of the difference $M_{\lambda_1=-5}-M_{\lambda_1=0}$. Here, in fact, the gradient of tissue $Q$ driving the haptotactic movement is bigger and, due to the negative value of $\lambda_1$, cells tend to avoid this area, moving away from it. Conversely, looking at the right plot of Figure \ref{Fiber_diff_lambda1}, we obtain exactly the reverse behaviour. In fact, the positive value of $\lambda_1$ leads to a much stronger haptotactic movement towards these fiber bundles. Thus, the difference shows positive values in the same regions described above.\\ 
\indent We then test the effect of varying the parameter $\lambda_0$, as described in experiment {\bf A.2}.  Results of this test are shown in Figure \ref{Diff_vary_lambda0}, where the difference between the solution of \eqref{macro_M_eq} for $\lambda_0=0.8\,(\text{s}^{-1})$ and the one obtained for $\lambda_0$ varying in the interval $[0.25,5] (\text{s}^{-1})$ are illustrated.

\begin{figure}[h!]
\centering
\includegraphics[width=0.85\textwidth]{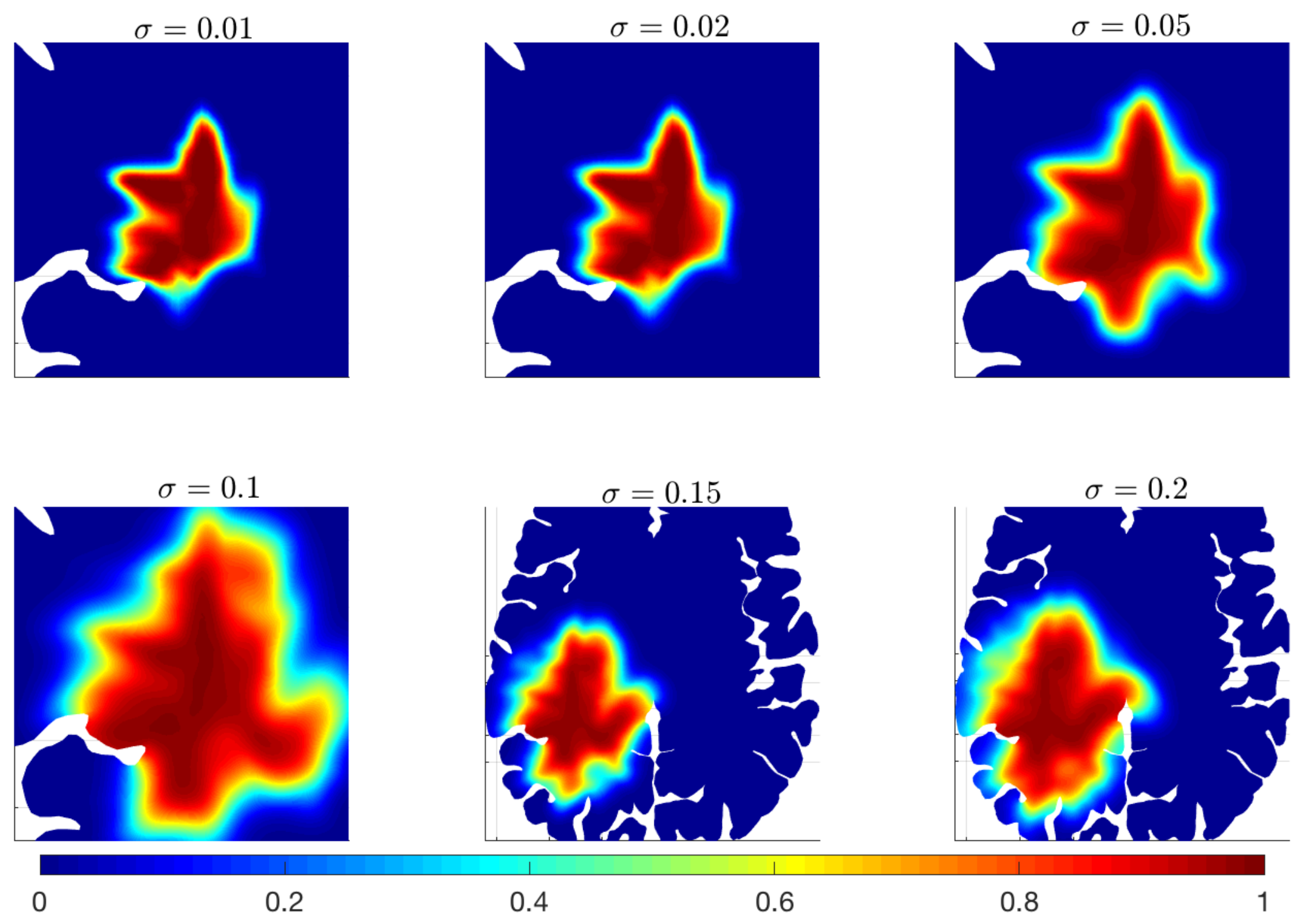}
\caption{{\bf Experiment (B).} Numerical simulations of equation (\ref{macro_M_eq}) with parameters listed in Table \ref{parameter_mod1} and for different values of $\sigma$. The tumour evolution is shown after 600 days. Values of $\sigma$ are expressed in mm$^2\cdot s^{-1}$. The figures referring to the cases $\sigma=0.15$ and $\sigma=0.2$ are shown on a less zoomed region to better assess the tumour invasion in the tissue.}
\label{Comp_sigma}
\end{figure}

\noindent We observe two different trends for $\lambda_0\ge0.8$ or $\lambda_0\le0.8$. Smaller values of the parameter lead to a larger spreading of the tumour cells with respect to the case $\lambda_0=0.8$, while larger values of it lead to a reduced invasion of the tumour mass. In fact, smaller values of $\lambda_0$ mean a reduced random turning of the cells, thus a greater persistence in their migration, which macroscopically translates into a large spread. Instead, larger values of $\lambda_0$ imply a larger frequency of cell turning and, thus, a macroscopic lower degree of persistence and spread in the tissue. In particular, the main difference is in the region of the outer rim of the neoplasia.\\
\indent Concerning the numerical test {\bf(B)}, we fix $\lambda_0=\lambda_1=0.8\,(\text{s}^{-1})$ and we vary the value of the parameter $\sigma$ relating to the variability of the cell velocity in the microscopic model \eqref{SDE_1} and, thus, leading the additional diffusion term appearing in the macroscopic model \eqref{macro_M_eq}. Results of the simulations for $\sigma \in [0.01-0.2]$ (mm$^2\cdot$s$^{-1}$) are shown in Figure \ref{Comp_sigma}.

\begin{figure}[h!]
\centering
\includegraphics[width=0.85\textwidth]{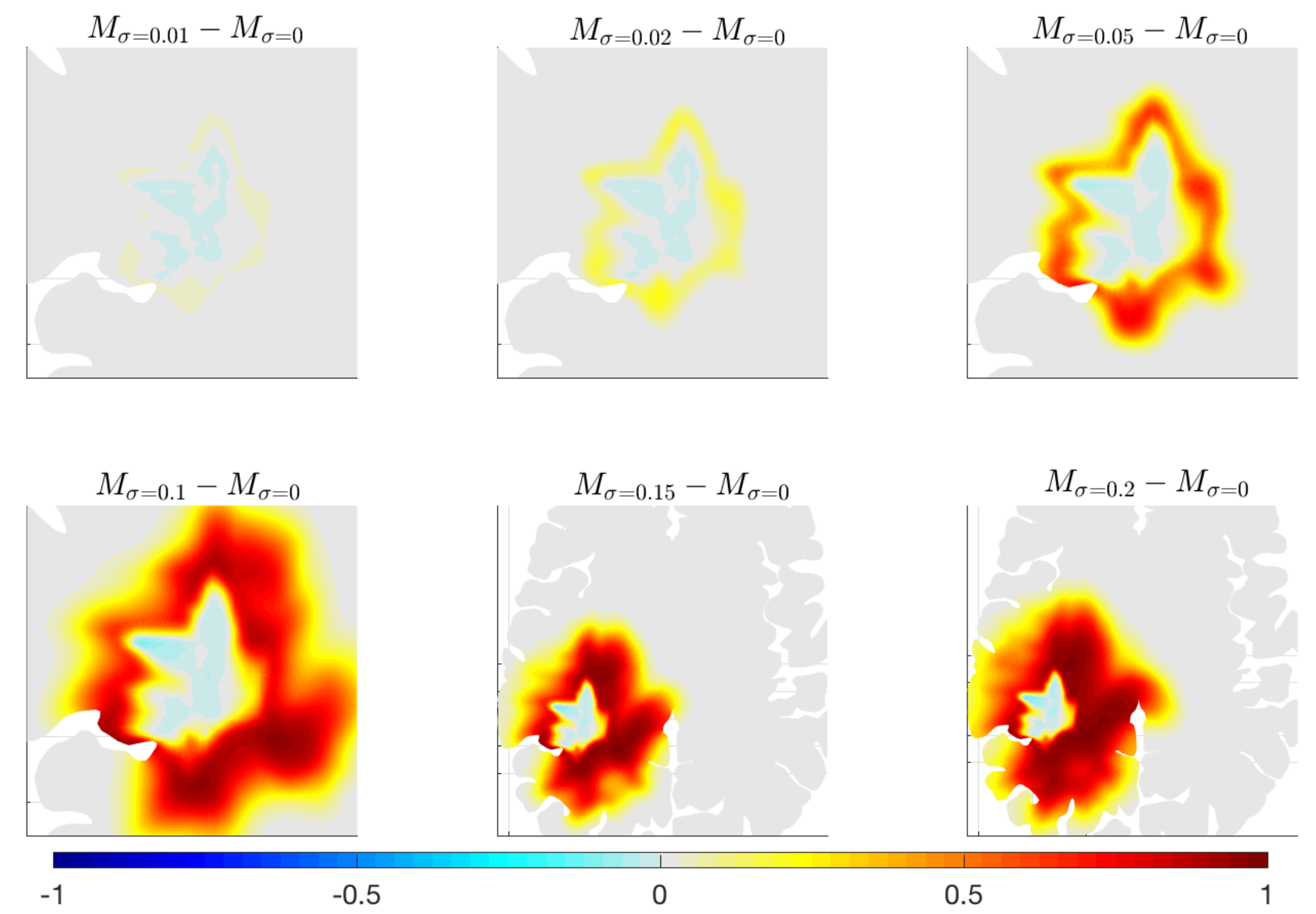}
\caption{{\bf Details of experiment (B).} Differences between the solution of system \eqref{macro_M_eq} for $\sigma=0$ and the one obtained for $\sigma\in[0.01,0.2]$ (mm$^2\cdot$ s$^{-1}$). Results are shown after 600 days. The remaining parameters are taken from Table \ref{parameter_mod1}. The figures referring to the cases $\sigma=0.15$ and $\sigma=0.2$ are shown on a less zoomed region to better observe the tumour invasion in the tissue.}
\label{Diff_sigma}
\end{figure}

\noindent As expected from equation \eqref{macro_M_eq}, the effect of the parameter $\sigma$ consists of a larger spread of the tumour cells inside the brain tissue. In particular, the larger the value of $\sigma$ is, the stronger the diffusion phenomenon characterising glioma cells appears. For large values of $\sigma$, we observe more regular tumour borders and a more isotropic cell migration because the additional diffusion term does not depend on the diffusion tensor $\eqref{DT_def}$. These features can be better appreciated in Figure \ref{Diff_sigma}, where the differences between the solution of equation \eqref{macro_M_eq} for $\sigma\ne0$ and for $\sigma=0$ are shown. 

\begin{figure}[h!]
\centering
\includegraphics[width=0.85\textwidth]{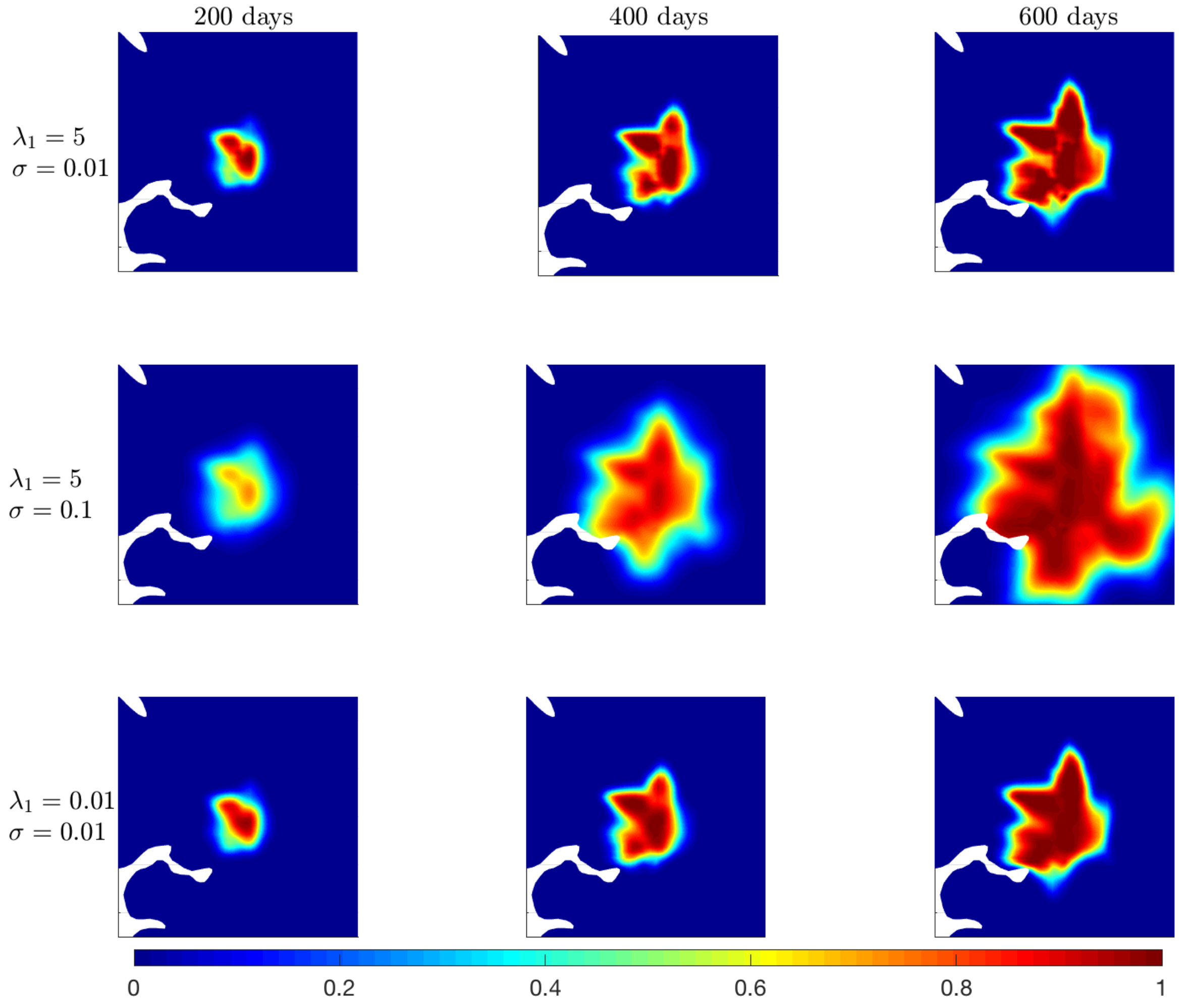}
\caption{{\bf Experiment (C).} Numerical simulations of equation (\ref{macro_M_eq}) with different combinations of $\lambda_1$ and $\sigma$. Columns refer to the three different time instants 200, 400, and 600 days, respectively. Rows refer to scenarios {\bf (C.1)}, {\bf (C.2)}, and {\bf (C.3)}, respectively. The remaining parameters are listed in Table \ref{parameter_mod1}. Values of $\lambda_1$ and $\sigma$ are expressed in s$^{-1}$ and mm$^2\cdot$s$^{-1}$, respectively.}
\label{Diff_scenarios_lambda_sigma}
\end{figure}

\noindent This figure clearly depicts an extensive and more homogeneous diffusion of the tumour mass for large values of $\sigma$. We obtain, in fact, negative values of the differences only in areas inside the tumour core (due to the balance between a faster spread and the same cell proliferation rate), while positive differences in the areas around the tumour border. In particular, comparing the first rows of Figures \ref{Diff_sigma} and \ref{Diff_vary_lambda0}, we notice that the increase of $\sigma$ values has an effect similar to the decrease of $\lambda_0$ values, i.e., a larger tumour spread in the area of tumour outer rim. It is interesting to observe how the same macroscopic cell behaviour is obtained from two  different microscopic processes. In fact, increasing $\sigma$ allows for a stronger effect of the stochastic component related to the variation of cell velocity, while decreasing $\lambda_0$ reduces the random turning of the cells and determines a greater persistence in their direction of migration. \\
\indent Referring to test {\bf (C)}, we analyse the interplay between the effects of the parameters $\lambda_1$ and $\sigma$. In particular, we consider three different combinations of them: 
\begin{itemize}
\item [{\bf (C.1)}] a high value of $\lambda_1$ and a small value of $\sigma$;
\item [{\bf (C.2)}] high values of both $\lambda_1$ and $\sigma$;
\item [{\bf (C.3)}] low values of both $\lambda_1$ and $\sigma$.
\end{itemize}
Results of these experiments are shown in Figure \ref{Diff_scenarios_lambda_sigma}.

\begin{figure}[h!]
\centering
\includegraphics[width=0.495\textwidth]{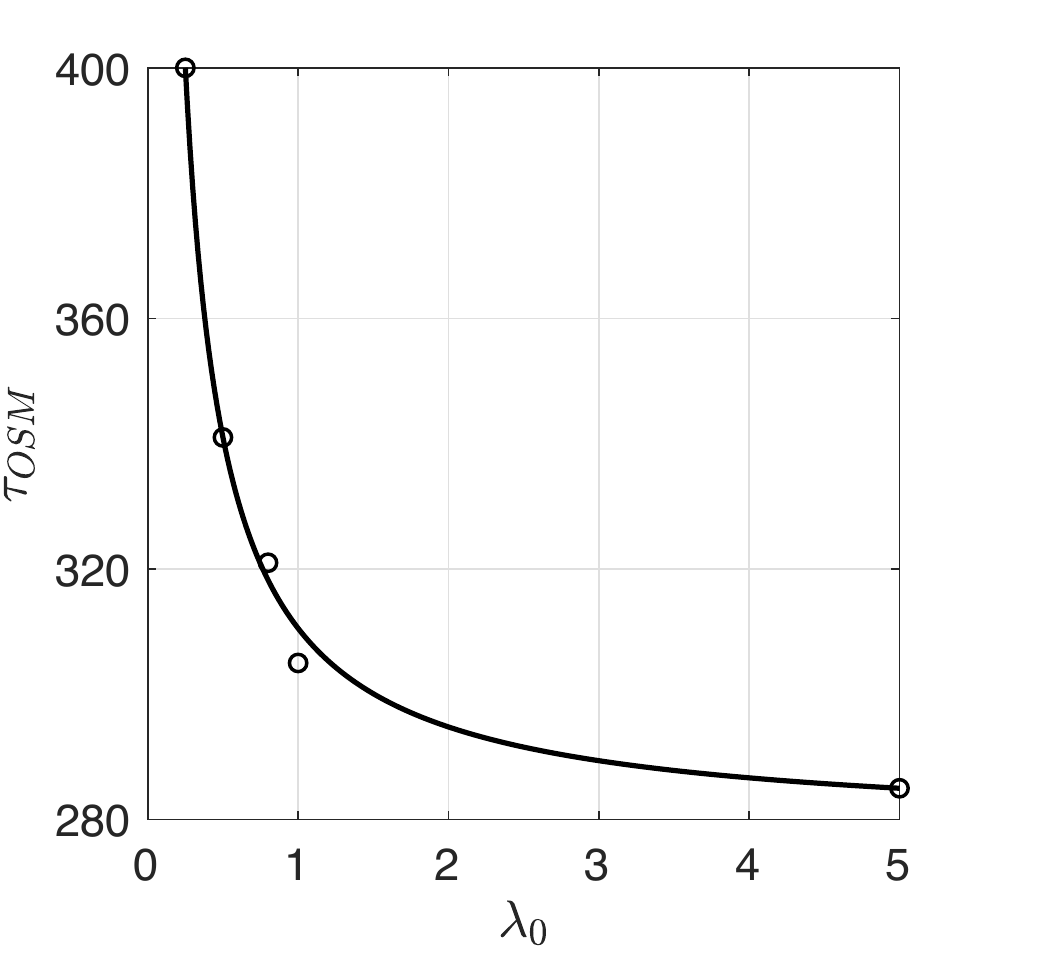}
\includegraphics[width=0.495\textwidth]{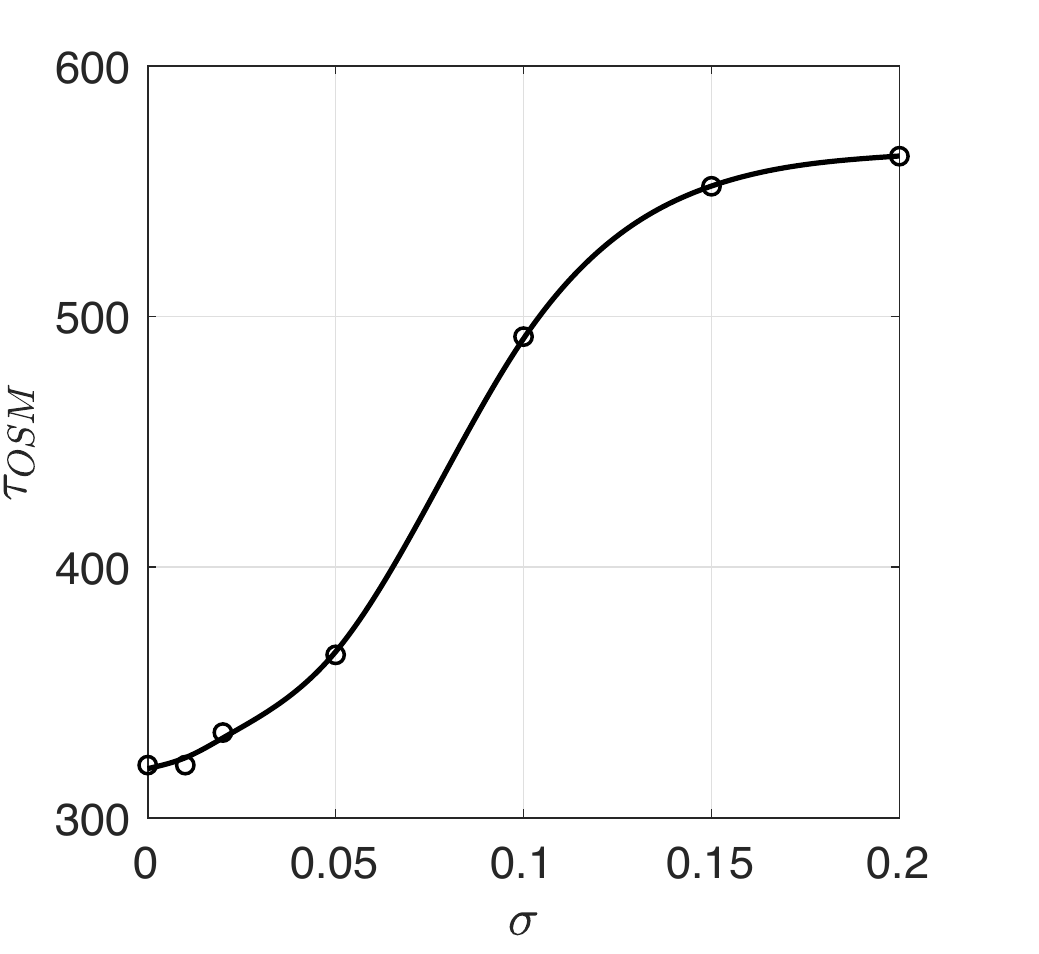}
\caption{{\bf Experiment (D).} Estimation of the time of onset of malignant transformation $\tau_{OSM}$ for different values of the turning rate $\lambda_0$ and the free parameter $\sigma$. The remaining parameters are taken as in Figure \ref{Evo_lambda1_08}.}
\label{T_osm}
\end{figure}

\noindent  From this figure, we notice how the respective effects of the variation of $\lambda_1$ and $\sigma$ (which we separately observed in the previous experiments {\bf (A.1)} and {\bf (B)} merge. In fact, in the scenario {\bf (C.1)} (first row of Figure \ref{Diff_scenarios_lambda_sigma}), the spread of the tumour cells is relatively confined due to the small value of $\sigma$. This spread follows the main fiber bundles present in the interested region, as $\lambda_1$ is large and drives the cell turning response to the fiber network. Moreover, the inner region of the tumour mass shows a high level of heterogeneity, as additional effect of the high value of $\lambda_1$. This heterogeneity becomes particularly evident comparing the tumour evolution at 600 days in the scenarios {\bf (C.1)} and {\bf (C.3)} (top and bottom row of Figure \ref{Diff_scenarios_lambda_sigma}), which use the same values of $\sigma$, but different values of $\lambda_1$. Considering the combination of high values for both parameters (scenario {\bf (C.2)}) leads to a larger spread of the tumour mass, as effect of the additional diffusion term driven by $\sigma$, and a different internal arrangement of the tumour cells compared with the bottom-left plot of Figure \ref{Comp_sigma} (where the tumour evolution is shown at $T=600$ days for $\sigma=0.1$ (mm$^2\cdot$ s$^{-1})$ and $\lambda_1=0.8$ (s$^{-1}$). This is still an effect of the higher value of $\lambda_1$, here set at $\lambda_1=5\,(\text{s}^{-1})$. Finally, the combination of low values for both $\sigma$ and $\lambda_1$ used for scenario {\bf (C.3)} determines a smoothness of the internal distribution of tumour cells as well as a reduced cell spread in the healthy tissue.\\
\indent For the last test {\bf (D)}, we discuss the onsets of malignant transformation from low grade glioma (LGG) to high grade glioma (HGG) in relation to the possible variations of the parameters $\lambda_0$, $\lambda_1$ and $\sigma$. LGGs are usually slowly-growing, infiltrative tumour with a very unpredictable clinical course. Most LGG patients face transformation of their tumour into higher grade one, with a worse prognosis. This process is known as malignant transformation and it is usually defined on the basis of contrast enhancement on MRI scans or histopathological evidences. In line with the approach proposed in \cite{bogdanska2017}, we estimate the time instant $\tau_{OSM}$ of the onset to the malignant transformation of cells into a more aggressive high grade tumour. The main aim of the proposed experiment {\bf (D)} consists in showing how our approach is able to replicate the same qualitative behaviours of \cite{bogdanska2017} (where a comparison with patient data is proposed), but with a more detailed and precise description of the microscopic processes related to cell migration. Specifically, $\tau_{OSM}$ is defined as the first time instant at which the LGG cell density becomes greater than a certain threshold $M_{crit}$, which we set to $0.6K_M$ \cite{bogdanska2017}. We run several numerical tests varying one parameter at the time and estimating the resulting time of onset of the malignancy. Table \ref{tau_est} collect the results of these experiments.
\begin{table} [!h]
\begin{center}
	\begin{tabular}{|c|c|c|c|c|c|c|c|c|c|c|} 
		\hline
		\rule{0pt}{2.5ex} $\lambda_1$ (s$^{-1}$) & -5 & -1 & -0.1 & -0.01 & 0 & 0.01 & 0.1 & 0.8 & 1 & 5\\
		\hline
		\rule{0pt}{2.5ex} $\tau_{OSM}$ (days)& 299 & 316 & 321 & 321 & 321 & 321 & 321 & 321 & 322 & 327\\
		\hline
	\end{tabular} \vspace{0.2cm}
\end{center}
\begin{center}
	\begin{tabular}{|c|c|c|c|c|c|c|c|c|c|c|} 
		\hline
		\rule{0pt}{2.5ex} $\sigma$ $(\text{mm}^2\cdot$ s$^{-1}$) & 0.01 & 0.02 & 0.05 & 0.1 & 0.15 & 0.2\\
		\hline
		\rule{0pt}{2.5ex} $\tau_{OSM}$ (days) & 321 & 334 & 365 & 492 & 552 & 564 \\
		\hline
	\end{tabular} \vspace{0.2cm}
\end{center}
\begin{center}
	\begin{tabular}{|c|c|c|c|c|c|c|c|c|c|c|} 
		\hline
		\rule{0pt}{2.5ex} $\lambda_0$ (s$^{-1}$) & 0.25 & 0.5 &0.8 & 1 & 5\\
		\hline
		\rule{0pt}{2.5ex} $\tau_{OSM}$ (days) & 400 & 341 & 321 & 305 & 285\\
		\hline
	\end{tabular} 
\end{center}
\vspace{0.3cm}
\caption{\footnotesize{{\textbf{Estimations of the onsets of malignant transformation $\tau_{OSM}$ for different values of $\lambda_1$, $\sigma$, and $\lambda_0$.}}}}
\label{tau_est}
\end{table}
\vspace{-0.3cm}
We observe that the parameter $\lambda_1$ seems to not have such an evident impact on the time of onset of malignancy. In fact, $\tau_{OSM}$ varies only of $\pm \,28$ days. Instead, both $\lambda_0$ and $\sigma$ strongly affect the estimation of $\tau_{OSM}$. In Figure \ref{T_osm} the estimated values of $\tau_{OSM}$ with respect to $\lambda_0$ and $\sigma$ are plotted together with the corresponding interpolant curves, showing the trends of $\tau_{OSM}=\tau_{OSM}(\lambda_0)$ (left plot of Figure \ref{T_osm}) and $\tau_{OSM}=\tau_{OSM}(\sigma)$ (right plot of Figure \ref{T_osm}).

\noindent Increasing the value of $\lambda_0$ leads to a reduction of the time $\tau_{OSM}$ at which LGG turns into HGG, while increasing $\sigma$ has the reverse effect, i.e., it leads to an increase of $\tau_{OSM}$. The parameter $\lambda_0$ is, in fact, related to the tumour responsiveness to the tissue structure, and large values of this parameter refer to a loss of responsiveness, which is a common characteristic in HGG. Moreover, observing that the overall diffusion coefficient of tumour cells in equation \eqref{macro_M_eq} is proportional to $\frac{1}{\lambda_0}+\frac{\sigma^2}{2}$, increasing $\sigma$ (or equivalently decreasing $\lambda_0$) corresponds to an increase of this diffusion coefficient. Thus, comparing these results with the ones shown in \cite{bogdanska2017} (e.g. see Figure 7 in there), we notice a good qualitative agreement between them and a similar behaviour for the evolution $\tau_{OSM}$. We would like to remark that this is only a first possible approximation for the estimation of $\tau_{OSM}$ and we are aware that there are several other factors involved in the definition of the transformation from LGG to HGG, apart from the increase in the tumour density. Surely, the tumour density values have an evident impact on the definition of $\tau_{OSM}$, however, from a mathematical point of view, it is difficult to provide a formal definition for it. Thus, as a first attempt, we decide to rely on the definition given in \cite{bogdanska2017} for $\tau_{OSM}$, leaving its possible extensions for future works. 

\section{Discussion}
\label{section5}
To the best of our knowledge, this is the first hierarchical stochastic model in which piecewise diffusion Markov processes are used to describe glioma cell motion within a multiscale framework. We start with the description of glioma cell movement at the microscopic scale using a PDifMP, which combines a stochastic model for cell motility and a deterministic one for cell migration. The latter looks at the response of glioma cells to external and environmental cues. The extended generator of the formulated PDifMP takes the form of an integro-differential equation in all the involved variables. Its solution yields the density of the transition probability of the Markov process. Using scaling arguments, we then obtain the equation describing the evolution of the tumor density at the macroscopic level. In this way, our approach allows us to take into account the macroscopic level properties as well as the features characterising the microscopic processes.\\ 
\indent Using numerical simulations of the macroscopic setting we analyse the role and influence of both the parameters involved in the jump rate function $\lambda$ of the PDifMP and the parameter $\sigma$ related to the stochastic variability in the cell velocity. In particular, we observe how the parameter $\lambda_0$ at the microscopic scale promotes a  major spreading of the tumour mass inside the brain tissue, regardless of the specific brain structure, while $\lambda_1$ relates to cell responsiveness to the guided movement along the brain fibers. The fully detailed formulation of glioma cell motion with the PDifMP allows us to observe that the jump rate function determines the distribution of the waiting times of the process being in a particular state. Thus, for a constant jump rate ($\lambda=\lambda_0$) there is no influence of the microenvironment on the motion and a larger frequency of cell turning determines at the macroscopic scale a reduced migration along the fibers. Instead, including the term $\lambda_1 z$ results in an increase in reorientations in response to the brain structure and, thus a visible heterogeneity inside the tumour bulk. A particularly interesting result is obtained by comparing the numerical experiments {\bf A.2} and {\bf B}. In fact, we show how a similar macroscopic behaviour - large cell spreading around the outer rim of the tumour -can result from two different sources at the microscopic level: either from increasing the value of $\sigma$ and thus the diffusion of cells, or from reducing the value of $\lambda_0$ and thus the random cell rotations, resulting in higher cell persistence .\\
\indent With respect to well-known multiscale models of this type \cite{engwer2015glioma,engwer2016multiscale,engwer2016effective}, in the present note we also include a further novel aspect concerning the transition to malignancy of the tumour mass. In particular, by accepting the hypothesis that the loss of responsiveness of glioma cells to the tissue structure can be seen as a sign of the transition from LGG to HGG, we numerically show that the time at which this transition happens can be estimated with our approach and it is highly influenced by the parameters $\sigma$ and $\lambda_0$. The obtained results are perfectly in line with the ones presented in \cite{bogdanska2017}, confirming the reliability of the proposed approach. \\ \indent With our work, we aim at emphasising how the use of PDMP or PDifMP for the description of the phenomena leading cell movement is of paramount importance for rigorously modelling the cellular scale processes. An interesting point would concern a numerical comparison of the cell behaviours at the different scales (microscopic and macroscopic) with either the deterministic and the stochastic formulation. Moreover, in the present notes, glioma cell motion is described in relation to the binding with the tissue, but the proposed approach can be extended in order to incorporate other biologically relevant aspects of tumour progression. For instance, following \cite{contesurulescu2021}, the influence of microenvironmental acidosis on glioma cell migration and the consequent pH-repellent chemotactic process can be considered. This could be done assuming different expressions for the jump rate function of the PDifMP, e.g. allowing its dependence on different interactions between cells and microenvironment or relating it to the tumour response to treatments. Another interesting direction for future development concerns the modification of the jump process using stochastic differential equations to model not only jumps in the cell velocity, but also jumps in the position, trying to recover the typical feature of tumour recurrence in different (and quite far from the original tumour location) regions of the brain. Finally, here we propose a first possible way to analyse the transition to malignancy. However, as stated in the above section, this process is much more complex and we are working towards the development of an interdisciplinary study in which an extension of our approach could be used to shed light on the intricate biological processes underlying this transition.

\begin{appendices}
\section{}
\subsection{Well-posedness of the macroscopic problem}
\label{Appendix_A}
\subsubsection{Assumptions}
Let $\mathbf{\Omega} \subset \mathbb{R}^3$ be a Lipschitz domain with a continuous boundary $\partial \mathbf{\Omega} \in C^{0,1}$ and let $\hat{n}$ be the normal vector to the boundary. Let $T>0$ such that $I=[0,T]$ denotes a finite time interval.
Let us define the Gelfand triple $(\mathcal{V},\mathcal{H},\mathcal{V}^*)$ (see \cite{gilbarg2015elliptic} for the definition) such that $\mathcal{V}=H^1_0(\mathbf{\Omega})=W^{1,2}_0(\mathbf{\Omega})$, $\mathcal{H}=L^2(\mathbf{\Omega})$, ${\mathcal{X}=L^2(I;\mathcal{V})}$, and $\mathcal{V}^*$ is the dual space of $\mathcal{V}$. We also define the functional space

$$\mathcal{W}:=\{ M \in \mathcal{X}=L^2(I;\mathcal{V}) : \partial_t M \in L^2(I;\mathcal{V}^* )\}$$
  such that $\mathcal{W} \dhookrightarrow \mathcal{X}$ (using conclusion 3.98 in \cite{ruuvzivcka2004fixpunktsatze} it is possible to prove the embedding).\\
  
\noindent We make the following assumptions.
	\label{assump_15}
	\begin{itemize}
		\item[A.1] The diffusion tensor $D_T(x)$ is positive definite, it belongs to the Sobolev space $W^{1,\infty}(\mathbf{\Omega})$ and its smallest eigenvalue is larger than a strictly positive constant $\alpha$.
		Note that $D_T(x)+\frac{1}{2}\sigma^2$ is also an element of $W^{1,\infty}(\mathbf{\Omega})$.
		\item[A.2] The function $\Gamma: \mathbb{R}\mapsto \mathbb{R}$ is continuous and it satisfies
		\begin{enumerate}
			\item the growth condition
			
			$$\vert \Gamma(s)\vert \leq c \big(1+\vert s\vert^{r-1} \big),$$
			where $c$ is a constant, independent from space and time, and $1\leq r < \infty$;
			\item the coercivity condition 
			
			$$\inf_{s\in \mathbb{R}_+}\Gamma(s)s>-\infty.$$
		\end{enumerate}
		\item[A.3] The function $Q(x)$ belongs to the Sobolev space $W^{1,\infty}(\mathbf{\Omega})$.
		\item[A.4] The velocity field $P_T(x)$ belongs to the Sobolev space $W^{1,\infty}(\mathbf{\Omega})$.
		\item[A.5] The term $\gamma(D_T(x),Q(x))$ belongs to the Lebesgue space $L^{\infty}$.
	\end{itemize}

\subsubsection{Existence}
\begin{theorem}
	Let $\Tilde{M}_0\in \mathcal{H}$ and the continuous function $\Gamma: \mathbb{R}\mapsto \mathbb{R}$ satisfies the condition $A.2$, with $1 \leq r \leq p\frac{2+n}{n}=\frac{10}{3}$, where $n$ denotes the \textsf{spatial dimension} ($n=3$) and $p$ refers to \textsf{the $p$-th power of the Lebesgue space} $L^p(\mathbf{\Omega})$ ($p=2$). Then, there exists a weak solution $M\in \mathcal{W}$, such that for all $\psi \in C^{\infty}_0([0,T]\times \mathbf{\Omega})$ it holds that
	
	\begin{equation*}
	\begin{split}
		&\int_0^T \langle \partial_tM,\psi \rangle_{\mathcal{V}}dt+\int_0^T\int_{\mathbf{\Omega}} ((D_T+\frac{1}{2}\sigma^2) \nabla M - \gamma(D_T,Q)M)\nabla \psi dxdt \\[0.1cm] 
		&+\int_0^T\int_{\mathbf{\Omega}} \Gamma(M) \psi(t) dxdt=0.
		\end{split}
	\end{equation*}
\end{theorem}
\vspace{-0.5cm}
\begin{proof}
	The proof is straightforward if we extend the proof proposed in \cite{engwer2016effective} to the case of the additional diffusion coefficient $\frac{1}{2}\sigma^2$.
\end{proof}
\subsubsection{Uniqueness}
\begin{proposition}
	Assuming that the function $\Gamma(M)$ is strictly monotone, the above solution to the macroscopic problem is unique.
\end{proposition}
\vspace{-0.5cm}
\begin{proof}
	See Lemma 3.38 and Theorem 3.66 in \cite{ruuvzivcka2004fixpunktsatze}, to prove the result. 
\end{proof}
\subsubsection{Non-negativity}
\begin{proposition}
	The solution of the macroscopic problem with $\tilde{M}_0\geq 0$ is non-negative.
\end{proposition}
\vspace{-0.5cm}
\begin{proof}
	We employ the truncation method to prove the non-negativity of the solution. Let $H(M)$ be a $C^{1,1}$ cutoff function such that:
	
	\begin{equation*}
		\left\{
		\begin{array}{ll}
			H(M(t,x)) & = \frac{1}{2} M^2 \quad M \in (-\infty,0),\\[0.2cm]
			H(M(t,x)) & = 0 \qquad\,\,\,\, M\in [0, \infty).
		\end{array}
		\right. 
	\end{equation*}
	We denote by $\phi(t)=\int_{\mathbf{\Omega}} H(\Bar{M(t,x)})dx$. We have:
	
	\begin{align*} \hspace{-0.5cm}
		\frac{d}{dt} \int_{\mathbf{\Omega}} H(\Bar{M}(t,x))dx=&\int_{\mathbf{\Omega}} H^{'}(M)\partial_t\Bar{M}(t,x)dx\\
		=& \int_{\mathbf{\Omega}} H^{'}(\Bar{M}\Big( \nabla_x.\big((D_T+\frac{1}{2}\sigma^2)\nabla_x\Bar{M}\big)-\nabla_x.\big(\gamma(D_T,Q)\Bar{M}\big)\\
		~& \, - \beta \Bar{M}+\Gamma(\Bar{M},Q)\Big)\\
		=& \int_{\mathbf{\Omega}} M\Big( \nabla_x.\big((D_T+\frac{1}{2}\sigma^2)\nabla_x\Bar{M}\big)-\nabla_x.\big(\gamma(D_T,Q)\Bar{M}\big)\\
		~&- \beta \Bar{M}+\Gamma(\Bar{M},Q)\Big)\,.
	\end{align*}
	We will now use the integration by part formula (1.69 in \cite{yagi2009abstract}  subsection 11.1) to get:
	
	\begin{align*}
	&	\int_{\mathbf{\Omega}} \Bar{M} \nabla_x.\big((D_T+\frac{1}{2}\sigma^2)\nabla_x\Bar{M}\big)=-\int_{\mathbf{\Omega}} \nabla_x\Bar{M} \big((D_T+\frac{1}{2}\sigma^2)\nabla_x\Bar{M}\big)dx\\[0.2cm]
		&\nabla_x.\int_{\mathbf{\Omega}} \big(\gamma(D_T,Q)\Bar{M}\big)dx= -\int_{\mathbf{\Omega}} \nabla_x\Bar{M} \big(\gamma(D_T,Q)\Bar{M}\big)dx.
	\end{align*}
	Hence,
	
	\begin{align*}
		&\frac{d}{dt} \int_{\mathbf{\Omega}} H(\Bar{M}(t,x))dx=-\int_{\mathbf{\Omega}} \nabla_x\Bar{M} \big((D_T+\frac{1}{2}\sigma^2)\nabla_x\Bar{M}\big)dx+\int_{\mathbf{\Omega}} \nabla_x\Bar{M} \big(\gamma\Bar{M}\big)dx \\[0.1cm]
		&-\beta\int_{\mathbf{\Omega}} \Bar{M}^2dx+\int_{\mathbf{\Omega}}\Bar{M}\Gamma(\Bar{M},Q)dx\\[0.1cm]
		&\leq \int_{\mathbf{\Omega}} \nabla_xM \big(\gamma\Bar{M}\big)dx+  \beta\int_{\mathbf{\Omega}} \Bar{M}^2dx+\int_{\mathbf{\Omega}}M\Gamma(\Bar{M},Q)dx \\
		&\leq \int_{\mathbf{\Omega}} \vert\nabla_x\Bar{M}\vert \vert\gamma\vert \vert\Bar{M}\vert dx +\beta\int_{\mathbf{\Omega}} \Bar{M}^2dx+\int_{\mathbf{\Omega}}\vert M\vert \vert\Gamma(\Bar{M},Q)\vert dx\\
		&\leq \int_{\mathbf{\Omega}} \vert \Bar{M}\vert\gamma\vert \vert\Bar{M}\vert dx +\beta\int_{\mathbf{\Omega}} \Bar{M}^2dx+\int_{\mathbf{\Omega}}\vert\Bar{M}\vert c(1+\vert\Bar{M}\vert^{r-1})dx\\
		&\leq \int_{\mathbf{\Omega}} \vert\nabla_x\Bar{M}\vert dx \int_{\mathbf{\Omega}}\vert\gamma\vert dx\int_{\mathbf{\Omega}}\vert\Bar{M}\vert dx+2\beta\int_{\mathbf{\Omega}} H(M)dx+\int_{\mathbf{\Omega}}\vert\Bar{M}\vert c(1+\vert\Bar{M}\vert^{r-1})dx,\\
		&\leq \int_{\mathbf{\Omega}} \vert\nabla_xM\vert^2dx \int_{\mathbf{\Omega}}\vert\gamma\vert dx\int_{\mathbf{\Omega}}\vert\Bar{M}\vert^2dx+2\beta\int_{\mathbf{\Omega}} H(\Bar{M})dx+\int_{\mathbf{\Omega}}\vert\Bar{M}\vert^2c(1+\vert\Bar{M}\vert^{r-1})dx,\\
		&\leq \Big(\Vert \nabla \Bar{M} \Vert^2_{L^2} \Vert \gamma\Vert_{L^{\infty}}+2\beta + C_1 \Vert \Bar{M} \Vert^r_{L^r}\Big)\int_{\mathbf{\Omega}} H(\Bar{M})dx.
	\end{align*}
	This implies that
	
	$$\frac{d}{dt}H(\Bar{M})\leq C_L H(\Bar{M})\,,$$
	Consequently, using Gronwall's inequality we have
	
	$$H(\Bar{M})\leq e^{C_Lt} H(0).$$
	Therefore, $\phi(0)=0$ implies that $\phi(t)=0$ for every $t>0$, namely, $\Bar{M}(t)\geq 0.$
\end{proof}
\end{appendices}

\section*{Acknowledgement}
E.B. and and A.M. were supported by the Austrian Science Fund (FWF): W1214-N15, project DK14, as well as by the strategic program ''Innovatives O\"O 2010 plus'' by the Upper Austrian Government. M.C. acknowledges funding by the Ministry of Education, Universities, and Research through the MIUR grant Dipartimento di Eccellenza 2018-2022, Project no. E11G18000350001, and the Scientific Research Programmes of Relevant National Interest project n. 2017KL4EF3. M.C. also acknowledges the support of the National Group of Mathematical Physics” (GNFM-INdAM). We thank Dr Philip-Rudolf Rauch from the University Hospital for Neurosurgery, Kepler University Hospital, for many fruitful discussions and insights into the topic of low grade gliomas.

\printbibliography

\section*{Declarations}
The authors have no competing interests to declare that are relevant to the content of this article.
\end{document}